\documentclass[prb,preprint]{revtex4-1} 

\usepackage{amsmath}  
\usepackage{amsfonts}  
\usepackage{graphicx}   
\usepackage{mathtools} 
\usepackage{enumerate}
\usepackage{subcaption} 
\usepackage{hyperref} 

\begin{document}

\title{Agent-based Monte Carlo simulations for reaction-diffusion models, population dynamics, and epidemic spreading}

\author{Mohamed Swailem}
\email{mswailem@vt.edu}
\affiliation{Department of Physics \& Center for Soft Matter and Biological Physics, 
    MC 0435, Robeson Hall, 850 West Campus Drive, Virginia Tech, Blacksburg, Virginia 24061, USA}
\affiliation{Laufer Center for Physical and Quantitative Biology, Stony Brook University, Stony Brook, New York 11794, USA}

\author{Ulrich Dobramysl}
\email{ulrich.dobramysl@gmail.com}
\affiliation{Peter Medawar Building for Pathogen Research, University of Oxford, Oxford OX1 3SY, U.K.}

\author{Ruslan I. Mukhamadiarov}
\email{ruslan.mukhamadiarov@gmail.com}
\affiliation{Arnold Sommerfeld Center for Theoretical Physics, Ludwig-Maximilians-Universit\"at M\"unchen,
	Theresienstr. 37, D-80333 Munich, Germany}

\author{Uwe C. T\"auber}
\email{tauber@vt.edu}
\affiliation{Department of Physics \& Center for Soft Matter and Biological Physics, MC 0435, Robeson Hall, 850 West Campus Drive, 
    Center for the Mathematics of Biosystems, Center for Emerging, Zoonotic, and Arthropod-borne Pathogens, Faculty of Health Sciences, 
    Virginia Tech, Blacksburg, Virginia 24061, USA}

\date{\today}

\begin{abstract}
We provide an overview of Monte Carlo algorithms based on Markovian stochastic dynamics of interacting and reacting many-particle systems not in thermal equilibrium.
These agent-based simulations are an effective way of introducing students to current research without requiring much prior knowledge or experience.
By starting from the direct visualization of the data, students can gain immediate insight into emerging macroscopic features of a complex system and subsequently apply more sophisticated data analysis to quantitatively characterize its rich dynamical properties, both in the stationary and transient regimes.
We utilize simulations of reaction-diffusion systems,  stochastic models for population dynamics and epidemic spreading, to exemplify how interdisciplinary computational research can be effectively utilized in bottom-up undergraduate and graduate education through learning by doing.
We also give helpful hints for the practical implementation of Monte Carlo algorithms, provide sample codes, explain some typical data analysis tools, and describe various potential error sources and pitfalls and tips for avoiding them.
\end{abstract}

\maketitle 

\section{Introduction} 
\label{introd}

The dynamics of many systems in nature incorporate interactions which induce a change of their fundamental constituents. 
Such reactive processes occur in chemistry, biochemistry, as well as in nuclear and elementary particle physics.
In addition, we may cast the kinetics of quasi-particles or other excitations in condensed matter systems in the language of chemical reactions. 
Examples include the mutual annihilation of topological defects with opposite winding numbers in continuum models,\cite{Nelson2002} the branching and fusion of structural or magnetic domain walls in 
solids,\cite{Krapivsky2010, Tauber2014} stochastic birth, death, and predation events in population dynamics,\cite{Haken1983, Murray2002, Krapivsky2010, Tauber2014} and the spreading of contagious diseases or opinions in social groups mediated through the agents' mutual contacts.\cite{Murray2002, Tauber2014}
Thus, reaction dynamics 
are important for modeling complex systems, and impact a broad range of basic and applied scientific problems.   
Successfully understanding these systems requires detailed experimental characterization and theoretical understanding through computational studies and mathematical analysis.

The fast development and wide availability of increasingly powerful computers have advanced the utilization of agent-based stochastic modeling or Monte Carlo simulations\cite{Newman1999, Binder2010, Gould2021} that can be implemented for well-mixed non-spatial systems as well as $d$-dimensional lattices and realistic networks.
The corresponding algorithms are based on  Markov chains, for which the stochastic time evolution of a system in the immediate future is completely determined by its present configuration and the prescribed transition rates into other configurations.
For a system either close to or relaxing toward thermal equilibrium, the stringent detailed balance condition severely restricts the possible choices of these transition rates.
In contrast, for nonequilibrium dynamical systems, no such general constraints apply, and we may consider the ensuing models to be determined by the set of configurations along with the associated Markovian transition rates connecting them.\cite{Kampen1981,Vliet2010,Krapivsky2010,Tauber2014}

The set of possible dynamical models away from thermal equilibrium is very rich.
Moreover, the absence of theoretical constraints eliminates the need for much prior knowledge and renders studies of non-equilibrium models by means of Monte Carlo simulations easily accessible to novice undergraduate and beginning graduate students.
The ensuing algorithms are straightforward to implement, parameters can be tuned to investigate the system's long-time and large-scale properties, and visual data representations such as movies for individual runs allow students to gain immediate insight into their dynamical behavior.
These observations may reveal unexpected emergent macroscopic features that include distinct and competing quasi-stationary or dynamical states, phase transitions and qualitative crossovers, dynamically induced correlations, and the related spontaneous formation of spatially inhomogeneous structures and/or temporal oscillations which may generate persistent traveling reaction fronts or activity waves pervading the system. 

Faculty can direct students' work toward scientifically relevant representations of natural phenomena, and provide direction on using appropriate time-series averages and their fluctuations, time and spatial Fourier transforms, and extracting static and dynamic correlations.
In this bottom-up manner, students are thus led from comparatively simple programs and parameter explorations to increasingly sophisticated data analysis.
In some cases students can be led to coarse-grained effective models with few degrees of freedom that are possibly amenable to analytical treatment or direct numerical solution, hence providing deeper insights into the nature of their observations.

Agent-based Monte Carlo simulations\cite{Durrett1999} can address important quantitative and qualitative effects and occasionally novel phenomena driven by random internal noise or external variability, and by dynamical correlations induced by the  kinetics. 
These effects are not typically treated by traditional textbooks on the dynamics of chemical, biological, ecological, and epidemiological systems.\cite{May1973, Maynard-Smith1974, Maynard-Smith1982, Haken1983, Hofbauer1998, Murray2002}
Investigations using coupled deterministic ordinary differential equations based on rate equations and mass-action type factorization of correlations are an often severe mean-field approximation.\cite{Tauber2014, Lindenberg2020, Tauber2020}
Near absorbing states characterized by the extinction of at least one species can lead to small particle numbers so that their discreteness becomes relevant and mean-field descriptions become irrelevant. 
Critical correlations may drastically amplify these fluctuations, implying strong quantitative effects on macroscopic observables\cite{Kuzovkov1988, Hinrichsen2000, Odor2004, Henkel2008, Tauber2014, Tauber2017, Lindenberg2020} not describable by deterministic rate equations.
Connections to percolation  may be used for their description,\cite{Stauffer1994, Janssen2005, Tauber2005} and established theoretical tools can be used to calculate fluctuation corrections and correlation effects.\cite{Goldenfeld1992, Cardy1996, Mattis1998, Tauber2012, Tauber2014, Serrao2017, Yao2023}
The spontaneous formation of spatio-temporal structures in nonequilibrium systems \cite{Cross1993, Schmittmann1998, Cross2009} may be considerably facilitated through stochastic fluctuations.\cite{Tauber2012, Tauber2014, Dobramysl2018, Yao2023, Tauber2024}
Over the past two decades our group has successfully engaged undergraduate and graduate students in investigations of fluctuation and correlation effects in diffusion-limited reactions,\cite{Deloubriere2002, Reid2003, Gopalakrishnan2004, Hilhorst2004a, Hilhorst2004b, Gopalakrishnan2005a, Gopalakrishnan2005b, Carroll2016, Deng2020, Tiani2024, Swailem2024}driven diffusive lattice gases,\cite{Daquila2011, Daquila2012, Mukhamadiarov2019, Mukhamadiarov2020, Nandi2021}, stochastic population dynamics,\cite{Mobilia2006, Mobilia2007, Washenberger2007, Dobramysl2008, He2010, He2011, He2012, Dobramysl2013a, Dobramysl2013b, Chen2016, Datla2017, Heiba2018, Chen2018, Serrao2021, Swailem2023} and epidemic spreading.\cite{Mukhamadiarov2021, Mukhamadiarov2022a, Mukhamadiarov2022b}

In  Sec.~\ref{meqsim}, we introduce the master equation for Markovian  processes as the underlying framework for Monte Carlo simulations and discuss the conceptual distinction between systems in and away from thermal equilibrium.
We then provide a straightforward  algorithm and sample code that can be readily modified for other applications and describe how typical observables, including correlation functions, are extracted from the data.
In Sec.~\ref{illexs}, we discuss examples  of the effects of reaction-induced spatial correlations in diffusion-limited pair annihilation reactions, the ensuing reaction fronts, and noise-generated spatio-temporal structures in simple predator-prey models.
We also briefly explore model variants that incorporate trait evolution through inheritance with random mutations, population extinction transitions, and address spatial and network extensions of paradigmatic models for the spreading of contagious diseases.
Section~\ref{tipcav} discusses tips and advice for practical   algorithm implementations, including relevant caveats and warnings that are not typically encountered in the research literature. We conclude with brief remarks    in Sec.~\ref{concls}.

\section{Monte Carlo simulations for stochastic reactions}
\label{meqsim}

\subsection{Theory of stochastic reaction processes: Master equation}
\label{stprme}

We will study local reactive processes that we shall model as random events governed by prescribed reaction rates.
To be concrete, let us consider the reversible binary reaction
\begin{equation}
	\mathrm{A + B} \xrightleftharpoons[\sigma]{\lambda} \mathrm{C},
\label{abcrea}
\end{equation}
with forward and backward  reaction rates $\lambda$ and $\sigma$.
In the limits $\lambda \to 0$ or $\sigma \to 0$, Eq.~(\ref{abcrea}) respectively reduces to the irreversible
first-order reaction $\mathrm{C} \xrightarrow{\sigma} \mathrm{A + B}$ and binary process $\mathrm{A + B} \xrightarrow{\lambda} \mathrm{C}$.
The left-hand side of Eq.~(\ref{abcrea})   implies that a particle of species $\mathrm{A}$ must meet a particle of species $\mathrm{B}$ at a specific location and time for a reaction to occur.

In a computer simulation, we frequently work with discretized space, for example, a regular hypercubic lattice in $d$ dimensions with characteristic spacing $b$, and with discrete time steps $\tau$. 
The reaction rates follow from assigned probabilities divided by $(\tau / N)$. 
For example, in the reaction in Eq.~(\ref{abcrea}) $\sigma \approx p_\sigma N_\mathrm{C} / \tau$, where $N_\mathrm{C}$ is the number of particles of species $\mathrm{C}$ with assigned probability $p_\sigma$. 
If a spatially continuous setting is implemented, we  allow reactions to occur within a small but finite radius $b > 0$. 
Such a cutoff is required, because point particles would otherwise encounter each other with vanishing probability.
We may also argue that on a sufficiently finely chosen mesh, at most only pair reactions can occur with a reasonable probability, and higher-order reactions are severely suppressed.
Consequently, we should view processes such as $k \mathrm{A} \to \ell \mathrm{A}$ with integers $k \geq 3$, $\ell \geq 0$ or $\mathrm{2 A + B \to C}$ as effective descriptions which become useful at length scales $|x|$ much bigger than $b$ and much smaller than the system's linear extension $L$.
In general, the relevant physics should be sought in this intermediate scale regime $b \ll |x| \ll L$, where a system's characteristic properties become independent of  model details and algorithmic constructs, and are also not influenced markedly by finite-size effects.

Markovian stochastic processes at time $t$ are fully determined by the system's immediately preceding configuration and the possible transitions with their accompanying rates to its new configuration.
These transition rates can be formally defined via conditional probabilities linking the initial and new states. A deterministic time evolution equation may be derived for the probability that a configuration $n$ is realized at time $t$.\cite{Kampen1981, Vliet2010, Tauber2014}
The ensuing  master equation  assumes the form of a balance between gain and loss terms, namely, probability flow respectively into and away from the state $n$ caused by transitions from and to all other states $\{ m \}$, $m \not= n$, governed by the positive rates $w_{n \to m} \geq 0$:
\begin{equation}
	\frac{\partial P_n(t)}{\partial t} = 
	\sum_{m \not= n} \Big[ P_m(t) w_{m \to n} - P_n(t) w_{n \to m} \Big] \ .
\label{masteq}
\end{equation}
The configurational probabilities $P_n(t)$ should remain normalized at all times, $\sum_n P_n(t) = 1$.
If we sum over $n$ in Eq.~(\ref{masteq}), we immediately satisfy normalization,  because we may exchange state labels $n \leftrightarrow m$ in the resulting double sum on the right-hand side so that the sum over the antisymmetric combination of terms in the square bracket yields zero:
\begin{equation}
	\frac{\partial}{\partial t} \sum_n P_n(t) = 
	\sum_{n, m \not= n} \Big[ P_m(t) w_{m \to n} - P_n(t) w_{n \to m} \Big]  = 0 \ .
\label{prcons}
\end{equation}

The transition rates $w$ usually depend on the configurations $\{ n \}$. 
However, if the transition rates do not carry an explicit time dependence, the coupled set of linear ordinary differential equations (\ref{masteq}) for the probabilities $P_n$ is closed, and the prescribed transition rates completely capture the evolution of the system. 
Any observable quantity $f$ is uniquely determined by the set of configurations $\{ n \}$, and its average value is given by
\begin{equation}
	\langle f(n,t) \rangle = \sum_n f(n) P_n(t) \ ,
\label{masexp}
\end{equation}
with its evolution determined by the time dependence of the configurational probabilities.
Thus, we may evaluate relevant averages, moments, and correlations to characterize the kinetics of the system.   

As an example, we construct the master equation for the stochastic reversible reaction in Eq.~(\ref{abcrea}).
These chemical processes encode the following stationary but configuration-dependent transition rates
\begin{subequations}
\begin{align}
	w_{\{ n_\mathrm{A}, n_\mathrm{B}, n_\mathrm{C} \} \to \{ n_\mathrm{A} - 1, 
	n_\mathrm{B} - 1, n_\mathrm{C} + 1 \}} & = \lambda \, n_\mathrm{A} n_\mathrm{B} \\
	w_{\{ n_\mathrm{A}, n_\mathrm{B}, n_\mathrm{C} \} \to \{ n_\mathrm{A} + 1, 
	n_\mathrm{B} + 1, n_\mathrm{C} - 1 \}} & = \sigma \, n_\mathrm{C} \ ,
\label{abctrr}
\end{align}
\end{subequations}
because  the forward and backward reactions with rates $\lambda$ and $\sigma$, respectively, have $n_\alpha$ possibilities to choose one of the reactants $\alpha = \mathrm{A,B,C}$.
Thus, the loss term for a configuration with $n_\mathrm{A}$, $n_\mathrm{B}$, and $n_\mathrm{C}$ particles at time $t$ is $- \left( \lambda \, n_\mathrm{A} n_\mathrm{B} + \sigma n_\mathrm{C} \right) P_{n_\mathrm{A}, n_\mathrm{B}, n_\mathrm{C}}(t)$.
For the gain processes from the forward reaction, we require $n_\mathrm{A} + 1$ and $ n_\mathrm{B} + 1$ reactant particles to be present initially, and there is at least one $\mathrm{C}$ particle in the final configuration. 
Similarly for the backward reaction, we must start with $n_\mathrm{C} + 1$ reactants, and $n_\mathrm{A}, n_\mathrm{B} \geq 1$.
With the helpful convention 
that $P_{\{ n_\alpha \}} = 0$ if any $n_\alpha < 0$, we arrive at the master equation that captures the stochastic processes in Eq.~(\ref{abcrea}),
\begin{eqnarray}
	\frac{\partial P_{n_\mathrm{A}, n_\mathrm{B}, n_\mathrm{C}}(t)}{\partial t} &=& 
	\lambda \left( n_\mathrm{A} + 1 \right) \left( n_\mathrm{B} + 1 \right) 
	P_{n_\mathrm{A}+1, n_\mathrm{B} + 1, n_\mathrm{C} - 1}(t) + \sigma \left( n_\mathrm{C}+1 
	\right) P_{n_\mathrm{A} - 1, n_\mathrm{B} - 1, n_\mathrm{C}+1}(t) \nonumber \\
	&&- \left( \lambda \, n_\mathrm{A} n_\mathrm{B} + \sigma n_\mathrm{C} \right) 
	P_{n_\mathrm{A}, n_\mathrm{B}, n_\mathrm{C}}(t) \ .
\label{abcmeq}
\end{eqnarray}
Because the particle numbers $n_\alpha = 0, 1, 2, \dots$ may take on any integer value, there is an infinite set of coupled differential equations.

The general master equation (\ref{masteq}) can be cast in matrix form as 
\begin{equation}
	\frac{\partial P_n(t)}{\partial t} = - \sum_m L_{nm} P_m(t)  \quad \textrm{with} \
	L_{nm} =  \delta_{nm} \sum_{m' \not= n} w_{n \to m'} - w_{m \to n \not= m} \ ,
\label{masmat}
\end{equation}
which formally reduces its solution to a standard linear algebra problem  in infinitely many dimensions.
For any (sub-)set of states that are connected by non-vanishing transition rates $w$, there consequently exists a unique and stable stationary configuration as $t \to \infty \,$. \cite{Kampen1981}
Because the Liouville matrix $L$ is not symmetric, its left and right eigenvectors differ, and its eigenvalues are in general complex numbers.
The eigenvalues' imaginary components represent the characteristic frequencies for the system's oscillatory dynamics, and their positive real parts yield the relaxation rates toward the stationary configuration.
Probability conservation is encoded in  $L$ by the constraint that its column sum must vanish: $\sum_n L_{nm} = 0$, which 
 implies the existence of at least one stationary left eigenvector with vanishing eigenvalue, namely, the projection state row vector $\left( 1, 1, \dots \right)$.

Observant readers will notice the formal analogy of the linear time evolution (\ref{masmat}) with quantum mechanics, but with real, positive probabilities constituting the fundamental time-dependent objects rather than complex wave functions. 
The generally non-symmetric Liouville matrix assumes the role of the Hermitian Hamiltonian operator to dictate the system's evolution, and the latter corresponding to a Schr\"odinger equation in imaginary time $i t / \hbar \to t$.
Nevertheless, due to the linear structure, we may construct appropriate Hilbert spaces and utilize well-developed mathematical techniques to characterize the properties of stochastic dynamics governed by master equations.
By using different basis representations, these master equations may assume apparently distinct forms that include a symplectic Hamiltonian structure\cite{Tauber2014} familiar from classical mechanics, which in this context is related to the generating function for statistical moments.\cite{Krapivsky2010}
We also mention the Doi--Peliti formalism, which exploits the fact that under any reactive process, particle numbers are changed by integer amounts.
Hence, we may invoke bosonic ladder operators familiar from the quantum harmonic oscillator to write $L$ in a non-Hermitian  form, and therefrom obtain a path-integral formulation that encapsulates the original stochastic master equation.\cite{Mattis1998, Tauber2005, Tauber2014}

\begin{itemize}

    \item[] {\bf Problem~1}: Construct the master equations for the
	\begin{enumerate}[(a)]
	\item   irreversible binary coagulation reaction $\mathrm{A + A} \xrightarrow{\lambda} \mathrm{A}$,
	\item Lotka--Volterra predator-prey competition model: $\mathrm{A} \xrightarrow{\mu} \emptyset$ (predator death).

\end{enumerate}
\end{itemize}

\subsection{Stationarity: global versus detailed balance and thermal equilibrium}

In the stationary regime, the configurational probabilities become time-independent, which implies that the net global probability current across all connected configurations, that is, the right-hand side of Eq.~(\ref{masteq}), vanishes.
A sufficient (but not necessary) condition for the absence of a global probability current is afforded by the cancellation of all individual terms in the square bracket for each state pair $\{ n, m \not= n\}$.
This  detailed-balance condition relates the stationary configurational probabilities with the forward and backward transition rates:
\begin{equation}
	\frac{P_n(\infty)}{P_m(\infty)} = \frac{w_{m \to n}}{w_{n \to m}} \ .
\label{detbal}
\end{equation}
This stringent set of conditions essentially defines the conditions for thermal equilibrium:
By concatenating Eq.~(\ref{detbal}) for an arbitrary configuration cycle $n_1 \to n_2 \to \ldots n_N = n_1$ of length $N-1$ and its reverse $n_N = n_1 \to n_{N-1} \to \ldots n_1$, the product of stationary probabilities on the left-hand side cancels, and we obtain the equivalent Kolmogorov criterion, which states that detailed balance holds if and only if
\begin{equation}
	w_{n_1 \to n_2} \, w_{n_2 \to n_3} \ldots w_{n_{N-1} \to n_1} = 
	w_{n_1 \to n_{N-1}} \, w_{n_{N-1} \to n_{N-2}} \ldots w_{n_2 \to n_1}
\label{kolmcr}
\end{equation}
for all possible cycles of arbitrary length.\cite{Kampen1981, Tauber2014}
Note that no information on the stationary probabilities $P_n(\infty)$ is required here.

The identities (\ref{kolmcr}) represent the complete absence of net stationary currents through any closed loops in configuration space.
The associated steady states are hence fully characterized by the stationary probabilities, which is the hallmark of thermal equilibrium.
For example, if micro-reversibility holds, $w_{n \to m} = w_{m \to n}$ for all $n \not=m$, we obtain the a priori equal steady-state probabilities $P_n(\infty) = P_m(\infty)$, which defines the microcanonical ensemble.
In contrast, for the canonical ensemble at fixed temperature $T$, the vanishing probability currents allow the characterization of the stationary states by a potential function $P_n = e^{\phi_n}$, where we may identify $\phi_n = - H_n / k_\mathrm{B} T$ with the Hamiltonian evaluated for configuration $n$.
The probability ratio in the detailed-balance condition (\ref{detbal}) then becomes potential or energy differences:
\begin{equation}
	\frac{w_{n \to m}}{w_{m \to n}} = e^{\phi_m - \phi_n} = e^{- (H_m - H_n) / k_\mathrm{B} T} \ .
\label{canens}
\end{equation}

For Monte Carlo simulations of a system in thermal equilibrium, the condition (\ref{canens}) constrains the choices of transition rates to ensure relaxation of the probabilities $P_n(t)$ to the canonical distribution.\cite{Newman1999, Binder2010, Gould2021}
A popular example is the Metropolis algorithm for which $w_{n \to m} = \mathrm{min}\left( 1 , e^{- (H_m - H_n) / k_\mathrm{B} T} \right)$, where
transitions from state $n$ to $m$ always occur when the energy is reduced ($H_m < H_n$), but are permitted with a nonzero probability if the energy increases, which allows the system to leave metastable configurations.

Note that the Kolmogorov criterion (\ref{kolmcr}) is violated if any of the processes involved are irreversible.
An example is an absorbing state, which can be accessed with finite rates, but for which no outgoing transitions are possible.
Population extinction is an example of such states, which cannot be described by equilibrium thermodynamics.
More generally, a complete quantitative characterization of nonequilibrium stationary states for which detailed balance does not hold requires the specification of the stationary probability currents in addition to the steady-state probabilities.\cite{Zia2006}
Conversely, if conditions (\ref{detbal}) or (\ref{kolmcr}) are valid for the set of transition rates even in an originally out-of-equilibrium setting, the system will relax toward a Boltzmann-like distribution with a possibly fictitious effective temperature.
Finally, if detailed balance holds, the Liouville matrix in Eq.~(\ref{masmat}) may be symmetrized by modifying the off-diagonal elements for $m \not= n$ according to \cite{Kampen1981,Tauber2014}
\begin{equation}
	L_{nm} \to {\widetilde L}_{nm} = - \sqrt{\frac{P_m(\infty)}{P_n(\infty)}} \, w_{m \to n} = 
	- \sqrt{\frac{P_n(\infty)}{P_m(\infty)}} \, w_{n \to m} \ .
\label{symlio}
\end{equation}
Consequently, right and left eigenstates become mere adjoints, and the eigenvalues are real.
This requirement implies that the approach to thermal equilibrium is purely relaxational, and no (damped) oscillatory kinetics is possible for the time-dependent probabilities $P_n(t)$.

\subsection{Ensemble averages and mean-field rate equations}
\label{mfrteq}

Equation~(\ref{masexp}) describes how time-dependent expectation values of any measurable quantity can be computed from the configurational probabilities $P_n(t)$.
We illustrate this process for the mean particle numbers for the reaction  in Eq.~(\ref{abcrea}) for  species $\mathrm{A}$:
\begin{equation}
	\frac{\partial \langle n_\mathrm{A}(t) \rangle}{\partial t} = 
	\sum_{n_\mathrm{A}, n_\mathrm{B}, n_\mathrm{C} = 0}^\infty n_\mathrm{A} \, 
	\frac{\partial P_{n_\mathrm{A}, n_\mathrm{B}, n_\mathrm{C}}(t)}{\partial t} \ .
\label{abcna1}
\end{equation}
We  need to substitute the master equation (\ref{abcmeq})  with appropriately specified summation limits, or impose the simpler convention that $P_{\{ n_\alpha \}} = 0$ if any $n_\alpha < 0$.
Inside the particle number sums that range over all allowed integer values, we may shift summation indices in the first gain term $\sim \lambda$ according to $n_\mathrm{A} \to n_\mathrm{A} - 1$, $n_\mathrm{B} \to n_\mathrm{B} - 1$, and $n_\mathrm{C} \to n_\mathrm{C} + 1$. Similarly, in the second gain term $\sim \sigma$,  $n_\mathrm{A} \to n_\mathrm{A} + 1$, $n_\mathrm{B} \to n_\mathrm{B} + 1$, and $n_\mathrm{C} \to n_\mathrm{C} - 1$.
These shifts yield
\begin{equation}
	\frac{\partial \langle n_\mathrm{A}(t) \rangle}{\partial t} = 
	\sum_{n_\mathrm{A}, n_\mathrm{B}, n_\mathrm{C}} \left[ 
	\lambda \left( n_\mathrm{A} - 1 \right) n_\mathrm{A} n_\mathrm{B}
	+ \sigma \left( n_\mathrm{A} + 1 \right) n_\mathrm{C} 
       - \lambda  n_\mathrm{A}^2 n_\mathrm{B} - \sigma  n_\mathrm{A} n_\mathrm{C} \right] 
	P_{n_\mathrm{A}, n_\mathrm{B}, n_\mathrm{C}}(t) \ ,
\end{equation}
which after cancelling terms in the square bracket allows us to see that the net overall reaction rate $R(t)$ is equal to the number of reactions per unit time in Eq.~(\ref{abcrea}) as a balance of certain time-dependent moments:
\begin{equation}
	\frac{\partial \langle n_\mathrm{A}(t) \rangle}{\partial t} = R(t) =
	- \lambda  \langle [n_\mathrm{A} n_\mathrm{B}](t) \rangle 
	+ \sigma  \langle n_\mathrm{C}(t) \rangle =   
	\frac{\partial \langle n_\mathrm{B}(t) \rangle}{\partial t} = 
	- \frac{\partial \langle n_\mathrm{C}(t) \rangle}{\partial t} \ .
\label{abcna2}
\end{equation}
The last two identities follow directly from the structure of   Eq.~(\ref{abcrea}). In general reaction processes,  the stoichiometric coefficients for the reactions would enter, which here are simply $1$.
In the steady state, as $t \to \infty$, the average particle numbers become constant and $R(\infty) = 0$, and we obtain the exact relation
\begin{equation}
	\frac{\langle n_\mathrm{C}(\infty) \rangle}{\langle [n_\mathrm{A} n_\mathrm{B}](\infty) \rangle} 
	= \frac{\lambda}{\sigma} \ .
\label{abcsts}
\end{equation}

To solve for the time-dependent or asymptotic particle species numbers in Eq.~(\ref{abcna2}), we  require the second moment $\langle n_\mathrm{A} n_\mathrm{B}(t) \rangle$, whose time derivative is determined by higher-order correlations (see Problem~3). Thus, the original set of infinitely many differential equations for the configurational probabilities $P_{n_\mathrm{A}, n_\mathrm{B}, n_\mathrm{C}}(t)$ becomes replaced by another infinite hierarchy of linear coupled differential equations for moments or correlations.

If we assume that sufficiently high-order correlations are insignificant and may be ignored to at least qualitatively capture the system's kinetics, we can achieve closure of the moment differential equations. 
The simplest and most stringent approximation is to neglect all two-point and higher correlations, which amounts to the factorization $\langle n_\mathrm{A} n_\mathrm{B}(t) \rangle \approx \langle n_\mathrm{A}(t) \rangle \langle n_\mathrm{B}(t) \rangle$. This approximation is reasonable for spatially well-mixed conditions or for reactions in dilute gases or solutions with abundant reactants where all spatial and time correlations are small relative to the mean densities.
In this mean-field approximation, the reaction rate in Eq.~(\ref{abcna2}) reduces to $R(t) \approx - \lambda \langle n_\mathrm{A}(t) \rangle \, \langle n_\mathrm{B}(t) \rangle + \sigma \langle n_\mathrm{C}(t) \rangle$, which yields a closed set of three coupled nonlinear ordinary differential equations for the mean particle numbers.

The mean-field factorization has thus drastically reduced the complexity of the original problem at the price of generating  nonlinear contributions in the rate equations.
These coupled ordinary differential equations can be analyzed by the tools of nonlinear dynamics. 
We may first identify their stationary states or fixed points, and then determine their stability and gain information about their basic dynamical properties via linearization about the steady-state configurations (see Problems~4 and 5).
We  note that the mean-field version of the steady-state result (\ref{abcsts}) becomes 
\begin{equation}
	\frac{\langle n_\mathrm{C}(\infty) \rangle}{\langle n_\mathrm{A}(\infty) \rangle \, 
	\langle n_\mathrm{B}(\infty) \rangle} = \frac{\lambda}{\sigma} \ ,
\label{abcstm}
\end{equation}
which is the chemical law of mass action,\cite{Haken1983, Lindenberg2020, Tauber2014} which is valid for reactions in dilute gases and solutions.
In thermal equilibrium, the reaction rate ratio $\lambda / \sigma = e^{-\Delta G / k_{\mathrm B}T}$ is determined by a reaction enthalpy $\Delta G$ and thus assumes the detailed-balance condition (\ref{canens}). 

\begin{itemize}

    \item[] {\bf Problem~2}: Derive the exact evolution equation for the mean particle number in binary coagulation $\mathrm{A + A} \xrightarrow{\lambda} \mathrm{A}$:
	$\partial \langle n_\mathrm{A}(t) \rangle / \partial t = 
	- \lambda \langle n_\mathrm{A}(t) (n_\mathrm{A}(t) - 1) \rangle$. 
	Then obtain the approximate mean-field rate equation and show that 
    $\langle n_\mathrm{A}(t) \rangle = \left[ \langle n_\mathrm{A}(0) \rangle^{-1} + \lambda \, t \right]^{-1}$.

    \item[] {\bf Problem~3}: Show that the evolution equation for the reaction scheme in Eq.~(\ref{abcrea}) is
	\begin{equation}
	\frac{\partial \langle n_\mathrm{A}(t) n_\mathrm{B}(t) \rangle}{\partial t} = 
	- \lambda \langle n_\mathrm{A}(t) n_\mathrm{B}(t) (n_\mathrm{A}(t) + n_\mathrm{B}(t) - 1)] \rangle 
	+ \sigma \langle (n_\mathrm{A}(t) + n_\mathrm{B}(t) + 1) n_\mathrm{C}(t) \rangle \ .
	\end{equation}
	Apply mean-field 	moment factorization and show that for large particle numbers $n_\mathrm{A}, n_\mathrm{B}, n_\mathrm{C} \gg 1$, the two-point correlation $\langle n_\mathrm{A}(t) n_\mathrm{B}(t) \rangle - \langle n_\mathrm{A}(t) \rangle \, \langle n_\mathrm{B}(t) \rangle$ remains approximately zero, provided it vanishes initially.

    \item[]  {\bf Problem~4}: Consider the Lotka--Volterra predator-prey  model with a finite prey carrying capacity $K$ as described by the (mean-field) rate equations for the mean predator ($\mathrm{A}$) and prey ($\mathrm{B}$)	populations:
    \begin{subequations}
	\begin{align}
    \frac{\partial \langle n_\mathrm{A}(t) \rangle}{\partial t} & = - \mu \langle n_\mathrm{A}(t) \rangle
	+ \lambda \langle n_\mathrm{A}(t) \rangle \langle n_\mathrm{B}(t) \rangle \ , \\
    \frac{\partial \langle n_\mathrm{B}(t) \rangle}{\partial t} & = \sigma \langle n_\mathrm{B}(t) \rangle
	\left[ 1 - \frac{\langle n_\mathrm{B}(t) \rangle}{K} \right] - \lambda \langle n_\mathrm{A}(t) \rangle
	\langle n_\mathrm{B}(t) \rangle \ .
 	\end{align}
	\end{subequations}
	Find and characterize the three stationary solutions for this system via the eigenvalues of the corresponding linear stability matrix.\cite{ArnoldODE1978}
  
    \item[] {\bf Problem~5}: The cyclic competition of three species is described by the stochastic ``rock-paper-scissors'' model with Lotka--Volterra type predation processes:\cite{Maynard-Smith1982}
	$\mathrm{A + B} \xrightarrow{k_{\mathrm A}} \mathrm{A + A}$, 
	$\mathrm{B + C} \xrightarrow{k_{\mathrm B}} \mathrm{B + B}$, 
	$\mathrm{A + C} \xrightarrow{k_{\mathrm C}} \mathrm{C + C}$.
	\begin{enumerate}[(a)]
	\item  Determine the corresponding coupled mean-field rate equations for the particle numbers and evaluate the associated linear stability matrix. \\
	\item  Find the stationary solution that describes species coexistence with mean particle numbers $\langle n_\mathrm{A}(\infty) \rangle, \langle n_\mathrm{B}(\infty) \rangle, \langle n_\mathrm{C}(\infty) \rangle > 0$. 
	Compute and interpret the stability matrix eigenvalues for this stationary coexistence solution.
\end{enumerate} 
\end{itemize}

\subsection{Markovian agent-based Monte Carlo simulations}
\label{mmcsim}

In agent-based Monte Carlo simulations of reaction processes and ecological or epidemiological kinetics, we implement the microscopic stochastic reaction processes on a computer as a Markov chain, where the updated configurations depend only on the immediately past states, and transitions between them are governed by prescribed configuration-dependent probabilities. 
Typically, spatially extended systems are simulated on regular  lattices, often employing periodic boundary conditions to minimize edge effects, or on  networks, which is especially relevant for a realistic modeling of epidemic spreading.
The underlying conceptual framework consists of stochastic master equations, although usually the simulations are not utilized to sample the time-dependent configurational probabilities, but aim at directly determining quantities such as particle numbers and/or their Fourier transforms and correlation functions.
Visualizing single runs is a very instructive means to obtain qualitative information about a system's evolution and provides useful hints as to which quantities should be employed to characterize its dynamical properties.

Any algorithm entails artifacts that affect its short-distance (on the scale of the lattice constant) and initial-time features.
These can be arbitrary, or determined by considerations of computational efficiency.
In any case, the goal is not to fully reflect a natural system's behavior on all scales, but to properly capture its emergent qualitative features at sufficiently long times and distances, including temporal oscillations or spatial structures.
These larger-scale features should be broadly independent of the algorithmic and simulation details,  if the system size is large enough for finite-size constraints to be ignored.

In most Monte Carlo runs, the simulation time is assumed to be proportional to ``real'' time. 
We  identify a Monte Carlo time step (MCS) as the number of random number calls in the program that would on average allow each particle in the system to have undergone a transition to another configuration.
The independence of observed meso- or macroscopic features from the algorithm details and microscopic details should be checked and confirmed.

\subsection{Sample simulation algorithm}
\label{ssalgo}

The agents in  agent-based simulations can range from indistinguishable passive particles belonging to a   species that are subject to certain reactive processes to truly active and perhaps  cognizant individuals who display a range of personal characteristics. 
We shall mostly be concerned with the former case, although we   briefly discuss agents that feature individual traits. 
We are concerned with spatially extended systems, meaning that particles are allowed to move or hop between sites on a lattice (or in a continuum).
Here, lattice means a next-neighbor connected arrangement of sites in $d$ spatial dimensions. 
However, there is no reason not to consider other arrangements, such as a randomly connected network of sites. 
Lattice sites may be constrained by a finite carrying capacity, which is a limit to how many particles are allowed to be present at a site at any  time.
This limit may range (aside from a zero or full exclusion) from a single particle to infinitely many particles.

Particles may interact with themselves or with other particles of their own species or another species. 
These mutual interactions can occur on the same lattice site at which the initiating particle is located, or they can occur between neighboring (or, usually less realistically, far distant) lattice sites. 
When lattice sites have a carrying capacity limited to a single particle, interactions necessarily have to be relegated to neighboring sites. 
The specifics of how these reactions or other interactions are determined by the problem and the researcher's goals. 

All of these choices, how much individuality agents have, the type of lattice, the carrying capacity, and the nature of the interactions determine the complexity of the algorithm which needs to be employed. 
To  begin, we introduce one of the simplest ways to simulate reaction-diffusion models and focus on the case of indistinguishable particles of one or more species, moving by nearest-neighbor hopping on a $d=2$ lattice with infinite carrying capacity. 
The algorithm we discuss advances the particles via a series of  steps consisting of a number of iterations. 
At the start of each iteration, a particle is selected at random from the pool of all particles comprising all involved species. 
The chosen particle then moves to a randomly chosen neighboring lattice site with a prescribed hopping probability $h$ (unbiased, if $h$ does not depend on the selected site and  biased if the transfer direction matters). 
Next, the particle is allowed to reproduce, interact with any other particles present on the same site, or perish in this order. 
All these interactions occur if they are relevant for the particle's species and proceed with pre-assigned probabilities. 
A  Monte Carlo step is complete when each particle that was present at the start of the step was selected once  on average. 

As an example, we consider the algorithm for the Lotka--Volterra predator-prey model, for which prey ($\mathrm{B}$) particles can reproduce with a probability $\sigma$, and predator ($\mathrm{A}$) particles can prey on species $\mathrm{B}$ with probability $\lambda$ and independently perish with probability $\mu$. 
Here, $\sigma$, $\lambda$, and $\mu$ represent the probabilities for a specific reaction to occur during one iteration of the algorithm. 

\begin{enumerate}

\item Select a particle $P$ at random regardless of species. Let $S$ be the species of particle $P$.

\item Let $E_P$ be the lattice site where $P$ resides, and $O$ be the set of sites adjacent to $E_P$. 
    Draw a random number $r_h \in [0,1]$. 
    If $r_h < h_S$,  randomly choose a member of $O$, where $h_S$ denotes the pre-set hopping probability of species $S$. 
    Then let $E_P$ be the new lattice site.

\item If $S = \mathrm{B}$, draw a random number $r_\sigma \in [0,1]$ and generate a new particle of species $\mathrm{B}$ at $E_P$ if $r_\sigma \leq \sigma$, the prey birth probability.

\item If $S = \mathrm{A}$, for each particle $U$ of species $\mathrm{B}$ on site $E_P$, generate a random number $r_\lambda \in [0,1]$. 
    If $r_\lambda \leq \lambda$, the fixed predation probability, remove particle $U$ and generate a new particle of species $\mathrm{A}$ at $E_P$.

\item If $S = \mathrm{A}$, generate a random number $r_\mu \in [0,1]$, and remove particle $P$ if $r_\mu < \mu$, the  predator death probability.

\end{enumerate}
These steps are performed $N_\mathrm{A}+N_\mathrm{B}$ 
times (i.e., the total number of $\mathrm{A}$ and $\mathrm{B}$ particles before the start of this Monte Carlo step) for one Monte Carlo step to be complete.

There is considerable freedom to choose the exact sequence of steps taken during an iteration. 
Microscopically, this choice matters.
For example, if step 5 came before step 4, a species $\mathrm{A}$ particle that has been selected to perish would not be able to prey on species $\mathrm{B}$, and thus would not be able to reproduce. 
As discussed in Sec.~\ref{mmcsim}, on a macroscopic level, this choice will typically imply a change of the effective reproduction rate of the predator species $\mathrm{A}$, but will not likely lead to a substantial change in the dynamics and qualitative outcome of the simulation. 
Similarly, allowing a species $\mathrm{A}$ particle to  prey only on a single $\mathrm{B}$ particle instead of each individual $B$ present on the same lattice site during the iteration would likely reduce the macroscopic observed predation rate and could  shift the region of coexistence of both species. 
It is therefore necessary to evaluate the order of rule execution and how different choices might influence the outcome in the context of a specific hypothesis under consideration. 

A  Python-based library accompanies this paper and can be downloaded from Ref.~\onlinecite{github1}. 

\begin{itemize}

 \item[] {\bf Problem~6}: The algorithm we have discussed is implemented as an interactive, browser-based Python Notebook.\cite{github2}   
    Familiarize yourself with the notebook and run the example without modifying it.

 \item[] {\bf Problem~7}: Swap the order of species $\mathrm{A}$'s {\tt PredationBirthReaction} and {\tt DeathReaction} in the example of Problem~6. 
    Run the simulation and observe how the time series for the species' densities changes.

\end{itemize}

\subsection{Typical observables}

Choosing the right set of observables to study a particular physical phenomenon  requires intuition that is gained by experience. 
One suggestion is to repeat the benchmark measurements for the system of interest before embarking on the search for potentially new physics. 
We may acquire a sense of the dynamics of a complex physical system by analyzing the same set of observables which are typically considered in the literature and by performing the same or similar measurements to check if a newly designed  code  reproduces known results at least qualitatively. 
In the following, we discuss a set of observables that are commonly used to study the stochastic dynamics and stationary properties of agent-based  simulations. 

A conceptually simple yet powerful key first step is the visual representation of the system's evolution. 
Such a movie  consists of a sequence of snapshots for a single simulation run and can help to verify the correctness of the code by direct inspection of important model features. 
For example, isotropic models should not display consistent anisotropic behavior; systems that should be homogeneous should not phase-separate; long-range order should be apparent when expected; systems that relax to equilibrium should not contain currents in the ensuing stationary state. 
We should  consider a few distinct and independent simulation runs to avoid basing inferences on a single unusual realization.
If everything appears in order, movies can then be used  to arrive at an initial understanding of the dynamical processes   in the system. 
We strongly recommend watching movies as the first analysis tool.
Based on the resulting better conceptual understanding, we can then select more atypical observables which may capture any observed new physics.
For example, if the system displays spontaneous phase separation in a particular direction, we may utilize the density profile to study how the height of the density kink changes with the control parameters. 
Or if the movies show marked clustering of agents, we could determine the associated correlation lengths and collect the resulting cluster size distribution. 
The information and understanding gained via this simple visual inspection should not be underestimated. 

A common quantity of interest for agent-based simulations is the overall density $n_\mathrm{A}(t)$ of each species as a function of time as measured in Monte Carlo steps. 
It characterizes the agents' global presence in the system at   time $t$, while the local density field $n_\mathrm{A}(\vec{x},t)$ provides additional information about their spatial distribution. 
The total density $n(t) = \sum_\mathrm{A} n_\mathrm{A}(t)$ can be useful for tracking the density of all species in the system (if this quantity is not conserved), as for example, in reaction-diffusion models and systems that are coupled to an external reservoir with which particles may be exchanged.
In continuous space, the calculation of the density in $d$-dimensions proceeds by assigning each particle a ``volume,'' typically a circle or (hyper)sphere with some small radius $b$. 
On a  lattice, the particle density of species $\mathrm{A}$ is computed by averaging all integer occupation numbers $n_\mathrm{A}(x,y,\ldots,x_d;t)$ over the system volume $V = L_x L_y \ldots L_d$,
\begin{equation}
    n_\mathrm{A}(t) = \frac{1}{L_x L_y \ldots L_d} \sum_{x}^{L_x} \sum_{y}^{L_y} \ldots \sum_{x_d}^{L_d} n_\mathrm{A}(x,y,\ldots,x_d;\,t) 
    = \frac{N_\mathrm{A}(t)}{V} \ .
\label{aveden}
\end{equation}
This spatial average should be supplemented with averages $\langle \ldots \rangle$ over sufficiently many independent runs, with different initializations and distinct time histories to reduce the statistical fluctuations to the desired level.
The  time-dependent mean densities $\langle n_\mathrm{A}(t) \rangle$ should then approximate the configurational ensemble averages (\ref{masexp}) whose evolution was discussed in Sec.~\ref{mfrteq}.

In contrast to the overall mean densities, the density fields $n_\mathrm{A}(\vec{x},t)$ provide local information about emerging inhomogeneities, such as clusters or the presence of spatially separated phases. 
Care must be taken in evaluating the ensemble average over independent simulation runs, because in statistically homogeneous and translation-invariant systems, the mean $\langle n_\mathrm{A}(\vec{x},t) \rangle$ becomes independent of the position $\vec{x}$ because the positions of clusters and phase boundaries are random for different simulation runs, and their contributions vanish when doing a simple ensemble average.
To avoid this problem, we compute the spatial Fourier transform 
\begin{equation}
    n_\mathrm{A}(\vec{q},t) = \! \int n_\mathrm{A}(\vec{x},t) \, e^{-i \vec{q} \cdot \vec{x}} d^dx = \sum_{\vec{x}} n_\mathrm{A}(\vec{x},t) \, e^{-i \vec{q} \cdot \vec{x}} = n_\mathrm{A}(-\vec{q},t)^* .
\label{spaftr}
\end{equation}
This quantity preserves information about the ensemble average of the spatial structure of the density.
Because the sum is only over occupied sites with $n_\mathrm{A}(\vec{x},t) = 1$, it can be written in terms of a sum over the sites $\vec{x}_i(t)$ occupied by the $N_\mathrm{A}(t)$ particles at time $t$: 
$n_\mathrm{A}(\vec{q},t) = \sum_{i = 1}^{N_\mathrm{A}(t)} e^{-i \vec{q} \cdot \vec{x}_i(t)}$, consistent with the  continuum density expression $n_\mathrm{A}(\vec{x},t) = \sum_{i = 1}^{N_\mathrm{A}(t)} \delta\big( \vec{x} - \vec{x}_i(t) \big)$.
Note that $n_\mathrm{A}(\vec{q}=0,t) = N_\mathrm{A}(t)$.
Fourier transforms are efficiently performed by using readily available Fast-Fourier-Transform packages.\cite{FFT}

If the system is in a steady state, that is, after sufficient time $t_0 \gg t_\mathrm{rel}$ beyond the characteristic relaxation time $t_\mathrm{rel}$ has elapsed, the resulting stationary mean occupation numbers can be computed by time averages,
\begin{equation}
    n_\mathrm{A}(\vec{x}) = \frac{1}{T - t_0} \sum_{t = t_0}^{T} n_\mathrm{A}(\vec{x};\,t) \ .
\label{locden}
\end{equation}
If the system has not yet reached its steady state, we can look for one or more directions in which the agent density stays constant, then average over particle occupation numbers along these  directions. 
A statistical average over different independent simulation runs  yields a well-defined oriented density profile. 
The resultant density field can reveal local density gradients, stationary or moving density shocks, phase-separated spatial regions, and more. 
The time Fourier transform
\begin{equation}
    n_\mathrm{A}(\omega) = \! \int n_\mathrm{A}(t) \, e^{i \omega t} dt
\label{temftr}
\end{equation}
can be used to extract model signatures. 
For example, in predator-prey populations the characteristic oscillation frequency may be obtained from the peak position of the Fourier signal $|n_\mathrm{A}(\omega)|$. 
The associated typical relaxation time scale can be measured by the peak's half-width at half maximum.\cite{James2011}

The local density fluctuation variance $\big(\Delta n_\mathrm{A}(\vec{x},t) \big)^2 = \big\langle \big( n_\mathrm{A}(\vec{x},t) - \big\langle n_\mathrm{A}(\vec{x},t) \big\rangle \big)^2 \big\rangle = \big\langle n_\mathrm{A}(\vec{x},t)^2 \big\rangle - \big\langle n_\mathrm{A}(\vec{x},t) \big\rangle^2$ in the bulk or at interfaces, obtained by averaging over independent  runs, can also provide useful information about the collective dynamics (see, for example, Ref.~\onlinecite{Schmittmann1998}). 
Similarly, local cross-correlations $\big\langle \big( n_\mathrm{A}(\vec{x},t) - \big\langle n_\mathrm{A}(\vec{x},t) \big\rangle \big) \big( n_\mathrm{B}(\vec{x},t) - \big\langle n_\mathrm{B}(\vec{x},t) \big\rangle \big) \big\rangle = \big\langle n_\mathrm{A}(\vec{x},t) \, n_\mathrm{B}(\vec{x},t)  \big\rangle - \big\langle n_\mathrm{A}(\vec{x},t) \big\rangle \big\langle n_\mathrm{B}(\vec{x},t) \big\rangle$ between distinct agent species can be illuminating.
In (statistically) homogeneous and stationary systems, these ensemble averages do not depend on $\vec{x}$ and $t$ because they are space- and time-translation invariant.
Hence, to investigate the spatial and temporal nature of fluctuations, we calculate the general dynamical density  correlation functions,
\begin{equation}
    C_{\alpha \alpha'}(\vec{x},t;\vec{x}\,',t') =
    \big\langle n_\alpha(\vec{x},t) \, n_{\alpha'}(\vec{x}\,',t') \big\rangle - \big\langle n_\alpha(\vec{x},t) \big\rangle \big\langle n_{\alpha'}(\vec{x}\,',t') \big\rangle \ ,
\label{dyncor}
\end{equation}
where $\alpha,\alpha' = \mathrm{A,B,}\ldots$ denote the species.
If spatial and temporal translation invariance holds, $C_{\alpha \alpha'}(\vec{x},t;\vec{x}\,',t') = C_{\alpha \alpha'}(\vec{x}-\vec{x}\,',t-t')$.

Evaluating correlation functions during the relaxation of a system provides the relevant length and time scales:
\begin{itemize}
\item At stationarity when $t \gg t_\mathrm{rel}$, we typically expect $C(\vec{x},0) \sim e^{-|\vec{x}| / \xi}$ at  distances $|\vec{x}| \gg \xi$, with $\xi$ the correlation length. The autocorrelations are also expected to decay exponentially in time as $C(0,t) \sim e^{-t / t_\mathrm{rel}}$.

\item In near-critical systems, the correlation length and relaxation time increase as  the power laws $\xi \sim \tau^{-\nu}$ and $t_\mathrm{rel} \sim \xi^z \sim \tau^{-z \nu}$, where  the parameter $\tau$ measures the distance to the critical point, and $\nu$ and $z$ are the correlation length and dynamical critical exponents, respectively.

\item At the critical point $\tau = 0$, the diverging correlation length and associated critical slowing down of relaxation processes induce algebraic behavior for the correlations, $C(\vec{x},0) \sim |\vec{x}|^{-(d-2+\eta)}$ and $C(0,t) \sim t^{- z (d-2+\eta)}$ with the Fisher exponent $\eta$.\cite{Goldenfeld1992, Cardy1996, Henkel2008, Tauber2014, Tauber2017}
\end{itemize}

Power law behavior for the correlation functions is also encountered in systems that exhibit generic scale invariance, such as driven lattice gases and, as we shall see, certain reacting particle models.\cite{Schmittmann1998, Tauber2014, Tauber2017}

\begin{figure}[t]
    \includegraphics[width=0.8\linewidth]{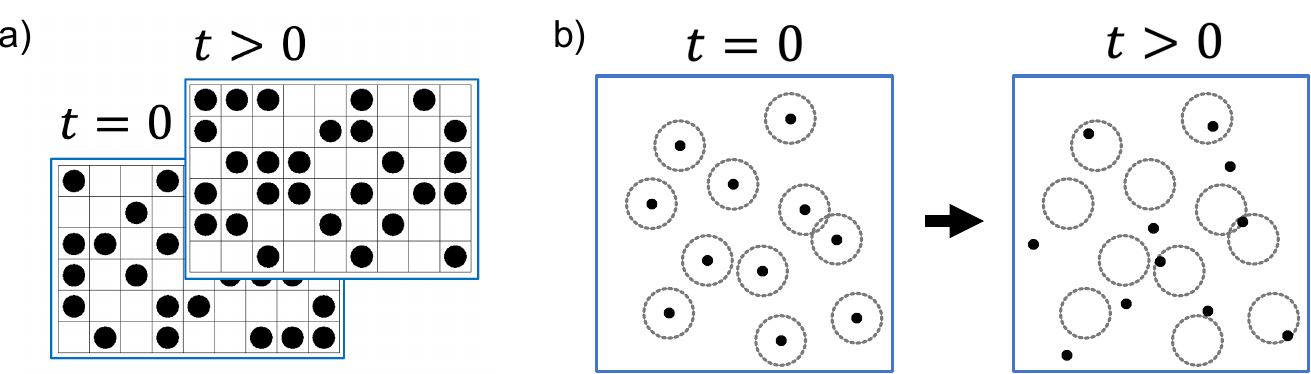}
    \caption{Calculation of the autocorrelation function $C(0,t)$  for (a) a lattice and (b) continuous space. 
    The autocorrelation function is a measure of how quickly a system forgets its initial state, which on a lattice is readily constructed by overlaying the snapshot at $t=0$ with one at  $t>0$. 
    In continuous space, we   establish a spherical volume of radius $b$ around each agent at $t=0$ which does not move with the agent at later times but remains fixed in space.
    The density autocorrelation function is computed by counting how many of these spheres contain particles at $t > 0$.}
    \label{fig:autorocc_func}
\end{figure}

On a lattice, the density autocorrelation function is given in terms of the occupation numbers at different times $t$ and $t'$ by [see Fig.~\ref{fig:autorocc_func}(a)]
\begin{equation}
    C_\mathrm{AA}(0;t,t') = \frac{1}{V} \sum_{\vec{x}} \big\langle n_\mathrm{A}(\vec{x},t) \, n_\mathrm{A}(\vec{x},t') \big\rangle - \langle n_\mathrm{A} \rangle^2 \ ,
\label{denaut}    
\end{equation}
To compute the autocorrelation function on a continuous space, we assign a sphere of radius $b$ to each agent, record the position of each particle at time $t=0$, and then track how many particles leave or re-enter the volumes prescribed by the initial particle position [see Fig.~\ref{fig:autorocc_func}(b)]. 
We often conveniently infer the structural characteristics of the system, such as its possibly emerging symmetries, spatial or temporal modulations, in Fourier space. 
For example, the Fourier transform of the density correlation function determines the coherent cross section in scattering experiments.
In a stationary, spatially translationally-invariant regime, the associated normalized dynamical structure factor is defined as
\begin{equation}
    S_\mathrm{AA}(\vec{q},t) = \frac{1}{N_\mathrm{A}} \big\langle n_\mathrm{A}(\vec{q},t) \, n_\mathrm{A}(-\vec{q},0) \big\rangle \ .
\label{strfac}    
\end{equation}
The static structure factor $S_\mathrm{AA}(\vec{q},0)$ is consequently  given in terms of the Fourier-transformed density (\ref{spaftr}), $S_\mathrm{AA}(\vec{q},0) = \big\langle |n_\mathrm{A}(\vec{q},0)|^2 \big\rangle / N_\mathrm{A}$.

We have not exhausted the list of useful observable quantities.
The general strategy of choosing an appropriate set of measurements is to first go over the relevant literature and look at available data. 
It takes much experience to develop a deeper intuition to identify the most relevant quantitative quantities for a given system to  capture its crucial dynamical features. 
We can train this intuition by taking the time to change parameters carefully and witnessing the ensuing change in several complementary observables.

\section{Illustrative examples}
\label{illexs}

\subsection{Spatial correlations in diffusion-limited annihilation reactions}
\label{dlannr}

{\em Annihilation reactions} serve as ideal introductory examples of reaction-diffusion models.
These models are conceptually simple to understand and straightforward to represent in a  simulation, and highlight how the dynamically emerging correlations and fluctuations can affect the ensuing kinetics. 
In general, annihilation models refer to a reaction between multiple particles of either the same or distinct species resulting in a decrease in the number of particles in the system.
The simplest nontrivial examples beyond spontaneous radioactive decay  are the single-species binary annihilation reaction $\mathrm{A} + \mathrm{A} \to \emptyset$, the single-species binary coagulation reaction $\mathrm{A} + \mathrm{A} \to \mathrm{A}$, and the two-species pair annihilation reaction $\mathrm{A} + \mathrm{B} \to \emptyset$. 
Particles may hop to adjacent lattice sites, or in the continuum limit spread diffusively. 
Although we focus on pair processes here, we can generalize these reactions to incorporate more input particles such as the general single-species annihilation reaction $k \mathrm{A} \to \ell \mathrm{A}$ with $0 \leq \ell < k$. 
We emphasize that each of these  irreversible annihilation reactions  violates the Kolmogorov criterion in Eq.~(\ref{kolmcr}). 
Our analysis for binary reactions should be sufficient to allow  readers to  investigate higher-order reactions with  $k \geq 3$ input particles. 

The single-species pair annihilation and coagulation reactions should qualitatively exhibit the same asymptotic long-time dynamics, even though the final absorbing states differ;  $N_\mathrm{A} = 1$ for the latter, and for the former either $N_\mathrm{A} = 0$ or $N_\mathrm{A} = 1$ if the initial particle number $N_\mathrm{A}(0)$ is even or odd, respectively. 
This assertion can be verified by Monte Carlo simulations and confirmed analytically by showing that the corresponding Liouville operator is the same (modulo numerical factors) in both models. 
Their qualitative equivalence can be rationalized by noting that the main difference between the two models is whether the number of $\mathrm{A}$ particles decreases by two or one when a reaction occurs; the qualitative behavior of the dynamics is actually determined by the conditions under which that reaction may occur, namely, that two particles must meet at the same site.

We begin by asking: What are the important physical features of this system? 
What are the relevant quantities that describe its properties? 
Starting with an initial configuration of reactants $\mathrm{A}$, they will annihilate until an absorbing state is reached, which by itself is not especially interesting.
Yet we can immediately conclude that the number density of $\mathrm{A}$ particles will monotonically decay to (essentially) zero. 
This behavior leaves just the question: How does the density decay? 
We will first describe a naive answer provided by the mean-field   approximation, and then discuss how emerging spatial anti-correlations  affect the system in low dimensions and alter the functional form of the density decay behavior.

\subsubsection{Single-species annihilation}
We first discuss the single-species annihilation (coagulation) model $2 \mathrm{A} \xrightarrow{\lambda} \mathrm{A}$ with reaction rate $\lambda$.
The exact  evolution for the mean particle number or density is (see Problem~2):
\begin{equation}
	\frac{\partial \langle n_\mathrm{A}(t) \rangle}{\partial t} = - \lambda \left\langle [ n_\mathrm{A} (n_\mathrm{A} - 1)](t) \right\rangle \ .
\label{eq:rate_eq_annihilation}
\end{equation}
In the absence of spatial correlations and for $N_\mathrm{A} \gg 1$, it may be approximated by the mean-field rate equation $\partial \langle n_\mathrm{A}(t)\rangle / \partial t = - \lambda \left\langle n_\mathrm{A}(t) \right\rangle^2$.
This ordinary differential equation is readily integrated and in the long-time  regime $\langle n_\mathrm{A}(0) \rangle \lambda t \gg 1$, we obtain the algebraic decay $\langle n_\mathrm{A}(t) \rangle \sim (\lambda t)^{-1}$, independent of the initial particle density $\langle n_\mathrm{A}(0) \rangle$ and the dimension $d$.

However, because the reactant density eventually decays to zero, the large-number assumption ceases to hold and spatially adjacent particles tend to annihilate each other with high probability.
Therefore, we anticipate that in the intermediate asymptotic regime, the density of $\mathrm{A}$ particles does not remain uniform, and the annihilation kinetics will generate large gaps between surviving particles. 
If these reaction-induced depletion zones that signify dynamical density anti-correlations cannot be filled effectively by diffusive mixing at sufficiently long times, the premise of the mean-field approximation must eventually break down. 
Hence, spatial anti-correlations and stochastic fluctuations ultimately determine the true asymptotic scaling behavior of the system, which is still described by a power law $\langle n_\mathrm{A}(t) \rangle \sim t^{- \alpha}$, but with a general decay exponent $\alpha \leq 1$.
We shall next explore this behavior using Monte Carlo simulations to determine the particle density decay in $d = 1$ and higher dimensions. 
After establishing the failure of the mean-field prediction for $d \leq 2$, we provide a heuristic argument that explains the observed density decays.

\subsubsection{Diffusion-limited scaling in one dimension}
\label{DLs1d}
The single-species binary annihilation or coagulation model $2 \mathrm{A} \xrightarrow{\lambda} \mathrm{A}$ may be readily simulated by means of the algorithm described in Sec.~\ref{ssalgo}, with the restriction that only a single particle is allowed on each lattice site.
(Allowing multiple site occupancies  introduces artifacts that will be discussed in Sec.~\ref{oalgoa}.)
We initiate the system with a completely filled lattice and with equal probabilities of attempting a coagulation reaction or a nearest-neighbor hopping process. 
The general situation of non-equal reaction and hopping probabilities will be addressed later.
To extract the density decay exponent $\alpha$, we need to compute the particle density in Eq.~(\ref{aveden}) as a function of time, which is easily obtained by an ensemble average from the Monte Carlo algorithm.

The functional dependence of the density on time is complicated. 
To focus on the long-time asymptotic scaling behavior, we write $\langle n_\mathrm{A}(t)\rangle$ in the form
\begin{equation}
	\langle n_\mathrm{A}(t)\rangle = A \, t^{-\alpha} + \epsilon(t) \ ,
\label{eq:density_scaling}    
\end{equation}
where $A$ denotes a constant amplitude, and $\epsilon(t)$ is a correction term that  becomes negligible compared to the leading power law for large times.
Instead of directly fitting the density to a power law, we first take the logarithm of Eq.~(\ref{eq:density_scaling}) to obtain:
\begin{subequations}
\label{eq:log_density_scaling}
\begin{align}
	\log \langle n_\mathrm{A}(t) \rangle &= \log\left[ A \, t^{-\alpha} + \epsilon(t) \right] = \log (A \, t^{-\alpha}) + \log\left[ 1 +\epsilon(t) / (A \, t^{-\alpha}) \right] \\
	&= \log A - \alpha \log t + \mathcal{O}\left( \epsilon(t) / (A \, t^{-\alpha}) \right) .
\end{align}
\end{subequations}
If we plot the logarithm of the density as a function of    $\log t$, we expect to  observe an approximately straight line, with an error term indicated in the second line.
Rather than attempting a direct power-law fit to the measured mean density, it is advisable to plot $\log \langle n_\mathrm{A}(t) \rangle$ as a function of $\log t$.
The negative slope of a linear fit to the long-time data yields the decay exponent $\alpha$.
 
\begin{figure}[t]
\begin{subfigure}[t]{0.48\textwidth}
    \includegraphics[width=\linewidth]{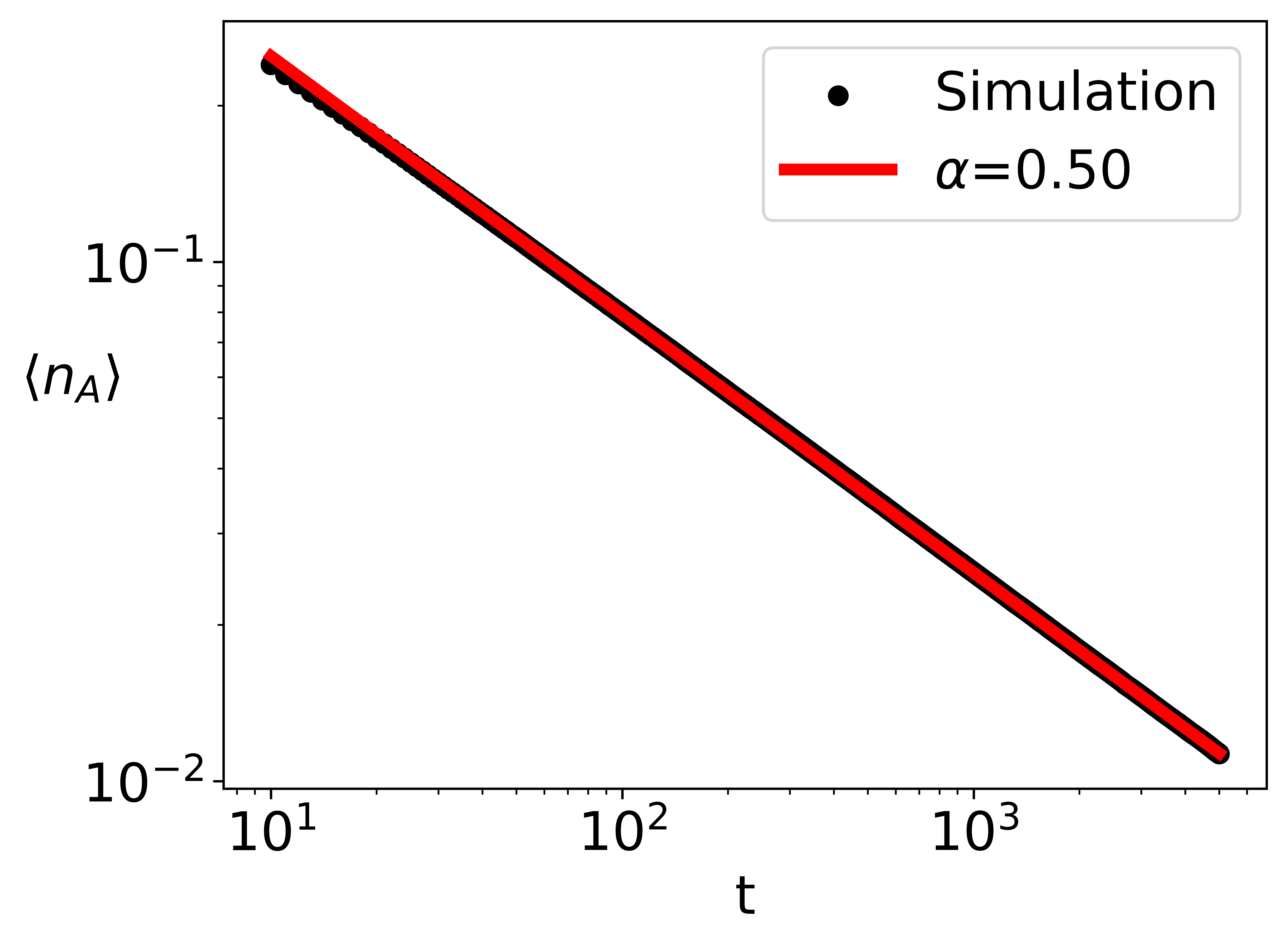}
    \caption{}
    \label{subfig:coagulation_density}
\end{subfigure}
\hfill
\begin{subfigure}[t]{0.48\textwidth}
    \includegraphics[width=\linewidth]{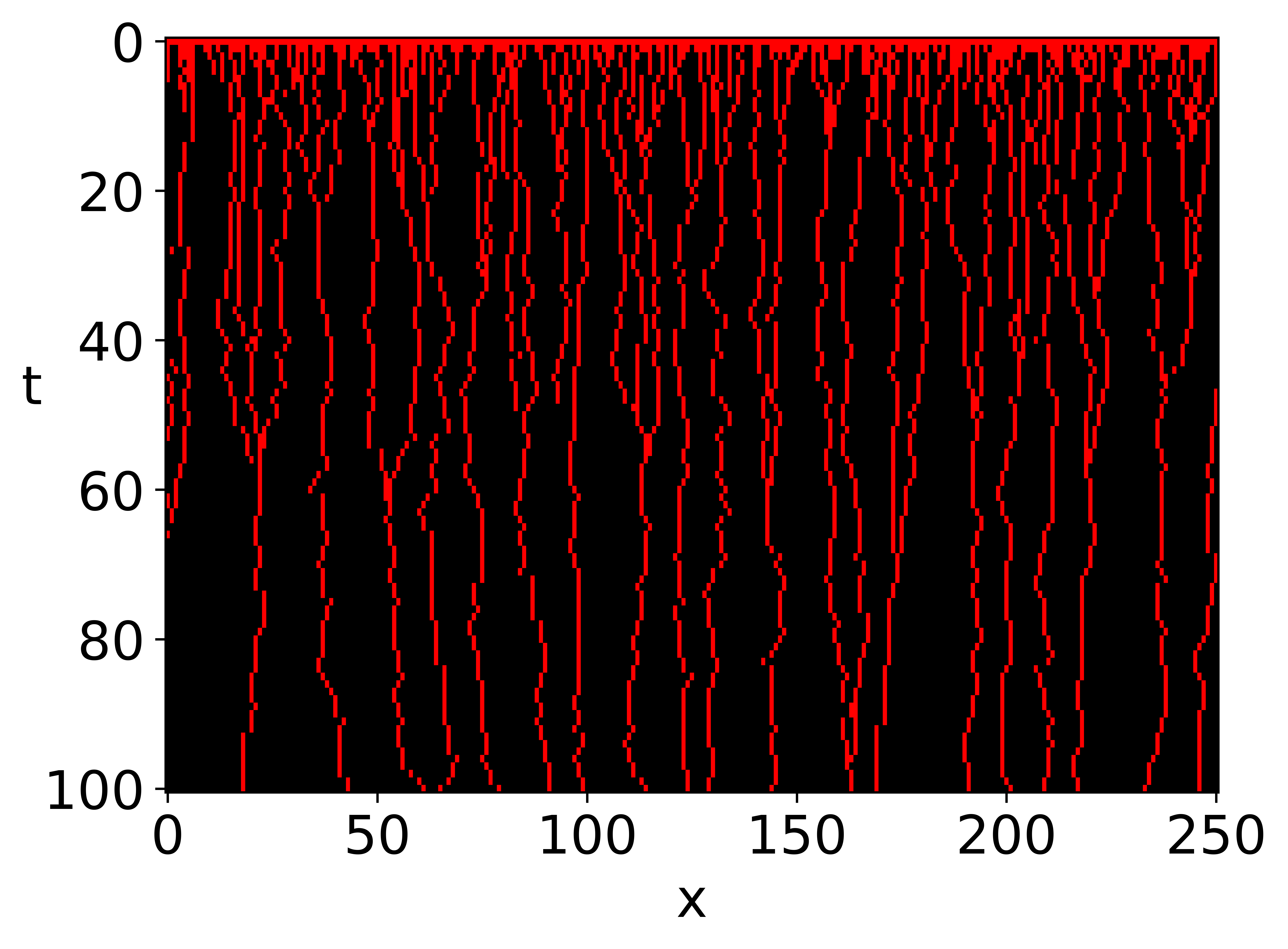}
    \caption{}
    \label{subfig:coagulation_snapshot}
\end{subfigure} 
    \caption{(a) Particle density decay as a function of time for the single-species binary annihilation model $2 \mathrm{A} \to \mathrm{A}$ in $d=1$. 
    The black line shows the  density measured in Monte Carlo steps of  lattice of length $L = 10^4$ with periodic boundary conditions. 
    The system was initially entirely filled, $N_\mathrm{A}(0) = L$, and equal annihilation and hopping probabilities $\lambda_0 = D_0 = 1$ were used. 
    The density data were averaged over $500$ independent  runs. 
    The red line indicates the result of fitting the simulation data to a straight line in the double-logarithmic plot. 
    (b) Simulation snapshot of a single realization displayed as particle ``world lines'' extending from top to bottom along the vertical time axis. 
    The horizontal axis represents the position along the chain; but the plot  features only the first $250$ lattice sites. 
    The presence of $\mathrm{A}$ particles is shown in red, and empty sites are shown as black.}
\end{figure}

We first focus on stochastic binary annihilation processes on a $d=1$ lattice with periodic boundary conditions, that is, a closed chain with $L$ sites.
Figure~\ref{subfig:coagulation_density} demonstrates that the measured density decay power law differs from the mean-field prediction, instead showing a much slower decay with exponent  $\alpha = 0.5$.
By inspecting the single  simulation displayed in Fig.~\ref{subfig:coagulation_snapshot}, it is evident that after a short transient reaction-limited time interval characterized by frequent annihilation reactions in the initial dense lattice, the system consists of isolated single particles that are clearly separated by slowly expanding depletion zones.
At this stage in the  dynamics, the chance of binary coagulation processes to occur is largely controlled by the time required for two reactants to reach each other.
This stage is called the diffusion-limited regime, because the effective reaction rate is set by the reactants'  mobility.
Diffusive spreading is parameterized by the diffusion constant $D$, 
which relates the typical distance travelled by any surviving particle to the elapsed time $t$: $\ell_D(t) \sim \sqrt{D t}$.

In the long-time regime, we may assume that the reaction rate $\lambda$ is ultimately large enough that once two particles meet, they also annihilate. 
Consequently any surviving particle must have already annihilated all other particles within the range $\ell_D(t)$.
Therefore, we expect that the linear size of the gaps between particles should be of the order of $\ell_D(t)$, giving an effective volume $\ell_D(t)^d$ per particle in $d$ dimensions.
This heuristic reasoning yields that the asymptotic particle density should scale as $\langle n_\mathrm{A}(t) \rangle \sim 1 / l_D(t)^d \sim (D t)^{-d/2}$, which is consistent with the simulation results for the $d=1$ lattice depicted in Fig.~\ref{subfig:coagulation_density}.
This diffusion-constrained density decay is slower than the reaction-limited $\langle n_\mathrm{A}(t) \rangle \sim (\lambda t)^{-1}$ for $d \leq 2$.
For $d > 2$, diffusive spreading effectively mixes the system, and the mean-field analysis is qualitatively adequate.
These arguments are a simplified version of the ingenious self-consistent analysis presented a century ago by Marian Smoluchowski (see Problem~8).\cite{Smoluchowski1916}

The reaction-limited regime is applicable when the diffusion rate is much larger than the reaction rate $\lambda$ so that the system remains well-mixed. 
In this case, the reactants find each other quickly, and their density decay is determined by the time taken for a reaction to occur once two particles meet. 
This regime recovers the mean-field decay law with $\alpha = 1$.
We anticipate that both scaling terms are present in the time dependence of the particle density, with each dominating at different times.
As a simplified ansatz, consider a decay function of the form $\langle n_\mathrm{A}(t) \rangle = A \, t^{-\alpha} + A' \, t^{-\alpha'}$, where $0 < \alpha < \alpha'$ are the scaling exponents that capture different regimes, and $A, A'$ are positive constants.
The long-time behavior of this function is dominated by the larger term $\sim t^{- \alpha}$. 
However, depending on the ratio of $A' / A$, this asymptotic limit might be observable only for extremely large times.
We expect the crossover to the long-time limit to occur at $t_\mathrm{cr}$ when the two terms become of the same order, that is, $A \, t_\mathrm{cr}^{-\alpha} \sim A' t_\mathrm{cr}^{- \alpha'}$, which gives $t_\mathrm{cr} \sim (A'/A)^{1 /(\alpha' - \alpha)}$, which may be very large if $A \ll A'$.
Therefore, we need to be careful in assessing the time behavior of such a system, because a large discrepancy in the prefactors of two competing power laws can push the asymptotic scaling limit beyond the accessible simulation time range.
This issue is discussed in more detail in Sec.~\ref{oalgoa}.

\begin{figure}[t]
    \includegraphics[width=0.7\linewidth]{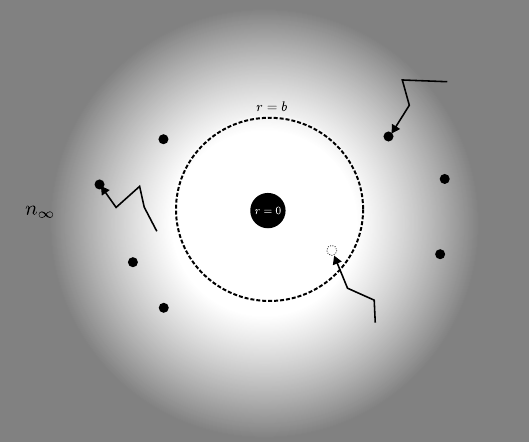}
	\caption{Sketch of the Smoluchowski model showing the density profile centered around the target particle located at the origin $r = 0$. 
    Arrows indicate the random relative diffusive motion of other reactants.     
    If any of these enter the reactive sphere of radius $b$, they coagulate with the center particle. 
    The resulting density gradient around the target particle captures a possibly emerging depletion zone; at large distances $r$, density fluctuations are averaged out to a constant asymptotic density value $n_\infty$.}
\label{fig:Problem1Fig}
\end{figure}

\noindent {\bf Problem~8}: 
    This problem systematically walks readers through Smoluchowski's treatment,\cite{Smoluchowski1916} which improves on the mean-field approximation by not assuming a uniform density, but constructs and solves a self-consistency equation to determine the emerging density profile. 
    
    We start by describing the system from the point of view of a single particle located at the origin and at a time already in the asymptotic region when depletion zones have developed. 
    We consider a continuum representation in which the particles propagate diffusively through $d$-dimensional space, subject to the diffusion equation $dn_\mathrm{A}({\vec x},t) / dt = D \nabla^2 n_\mathrm{A}(\vec{x},t)$, and the annihilation reaction occurs with probability one when two particles meet. 
    In a continuous system, the probability of two particles being in the same location is zero, and it is necessary to introduce a reactive sphere of radius $b$ around each particle. 
    If the distance between two particles is less than $b$, then the annihilation reaction occurs (Fig.~\ref{fig:Problem1Fig}). 
    Following Smoluchowski, we assume that in the asymptotic regime, reactions rarely occur due to the large gaps between particles. 
    Therefore the rate of density change is much slower than the underlying diffusion process, which implies that we can assume the density profile reaches a quasi-steady state governed by the stationary diffusion (Laplace) equation $\nabla^2 n_\mathrm{st}(r) = 0$.
    
    \begin{enumerate}[(a)]
    \item Obtain the steady-state diffusion equation in spherical coordinates in $d$ dimensions, and impose spherical symmetry  to simplify it. 
    \item Solve the steady-steady diffusion equation to obtain $n_\mathrm{st}(r)$. 
    Your answer should include two integration constants $C_1$ and $C_2$. 
    \item To determine $C_1$ and $C_2$, we need to impose constraints on the density that are guided by  intuition. 
    Because particles that enter the reactive sphere are annihilated, no other reactants should reside within it: $n_\mathrm{st}(r = b) = 0$. 
    The reactants' influence should be local, and hence the particle density should average to a constant background value far  from the target particle, $n_\mathrm{st}(r \to \infty) = n_\infty$. 
    Use your solution and show that these boundary conditions are  applicable only for $d > 2$. 
    The value $d_c = 2$ is known as the upper critical dimension for stochastic binary annihilation with diffusive spreading. 
    Apply these boundary conditions for $d > 2$ to obtain the desired solution.
    \item With the spatial density profile thus established, we can relate it to the  evolution of the density by computing the steady-state particle influx into the reaction sphere, normalized by the total particle number.
    If we assume that all inflowing reactants are annihilated, we can directly identify this quantity with the effective reaction rate $\lambda_\mathrm{eff}$. 
    Because we affixed the target particle to the origin, other reactants propagate relative to it with diffusivity $2 D$. 
    The total flux across the surface is then $\Phi = 2 D \! \int_{\partial b} \vec{\nabla}n_\mathrm{st} \cdot d\vec{S}$, where $\partial b$ denotes the boundary surface of the reactive sphere of radius $b$. 
    For $d > 2$, use this prescription to compute the total flux and relate it to the effective reaction rate $\lambda_\mathrm{eff} \sim D$.
    Show that the result is the mean-field decay exponent $\alpha = 1$, with a modified amplitude that   vanishes as $d \to 2$.
    \item For $d < 2$, we cannot apply the same boundary condition for $r \to \infty$, because the solution diverges in that limit. 
    To treat this divergence, utilize the emerging characteristic spacing $R$ between the particles to define a typical length scale beyond which the density becomes approximately homogeneous.     Because the depletion zone volumes $\sim R^d$ grow with time as the density decreases, the typical spacing between particles depends on $n_\mathrm{st}$ in a way that is determined self-consistently. 
    Quantitatively, we have on average $n_\mathrm{st}(R) = 1 / (K_d R^d / d)$, where $K_d$ is the surface area of a $d$-dimensional unit hypersphere ($K_2 = 2 \pi$, $K_3 = 4 \pi, \ldots$), so $R(n_\mathrm{st}) = \left( n_\mathrm{st} K_d / d \right)^{-1 / d}$.
    Apply the modified boundary condition $n_\mathrm{st}(R) = n_\infty$, with a quasi-stationary fixed $R$, to obtain the solution $n_\mathrm{st}$ as a function of $R \gg b$, assuming that the depletion zones are much larger than the microscopic  cutoff $b$ in the long-time limit and keeping only the zeroth-order term in a $b/R$ expansion. 
    The final (approximate) solution should be $n_\mathrm{st}(r) = n_\infty \left( r / R(n_\mathrm{st}) \right)^{2-d}$.

\item Obtain the normalized particle influx at $r = b$ for $d < 2$  as in part~(d) for fixed $R$. 
    The resulting effective reaction rate $\lambda_\mathrm{eff}(n_\mathrm{st})$ looks similar to the case for $d > 2$, with the factor $d - 2$ and the cutoff $b$ replaced by $2 - d$ and the density-dependent  cutoff $R(n_\mathrm{st})$, respectively. 
    Substitute $\lambda_{\text{eff}}(n)$ into the mean-field rate equation and obtain the ensuing particle density decay power law.
    \item The situation at the critical dimension $d=d_c = 2$ is   more subtle, because the solution to the diffusion equation involves logarithmic terms. 
    Solve the stationary diffusion equation specifically for $d = 2$, with appropriate boundary conditions to find the normalized particle influx at the reaction zone boundary $b$ and relate it to the effective reaction rate $\lambda_{\text{eff}}$. 
    Rather than directly integrating the corresponding rate equation, express the solution in the  iterative form $n(t) \sim 1 / \int_0^t \lambda_\mathrm{eff}\big( n(t') \big) dt'$, and assume to leading order that $n$ on the right-hand side is constant.
    Simplify the final answer for $\ln (D t) \gg 1$.
    \end{enumerate}

\subsubsection{Higher dimensions}
We have shown that  diffusion-limited scaling dominates  pair annihilation or coagulation density decay in  $d=1$.
As  mentioned,  diffusion-limited decay becomes faster with increasing $d$, and the  decay exponent $\alpha = d/2$ reaches its reaction-limited or mean-field value $\alpha=1$ at the upper critical dimension $d_c = 2$.
As discussed in Problem~8(g), there are logarithmic corrections in  $d=2$  to the mean-field power scaling behavior, which is a general feature of critical phenomena at their upper critical dimension.\cite{Goldenfeld1992, Cardy1996, Hinrichsen2000, Henkel2008, Tauber2014}

\begin{figure}[t]
\begin{subfigure}[t]{0.48\textwidth}
    \includegraphics[width=\linewidth]{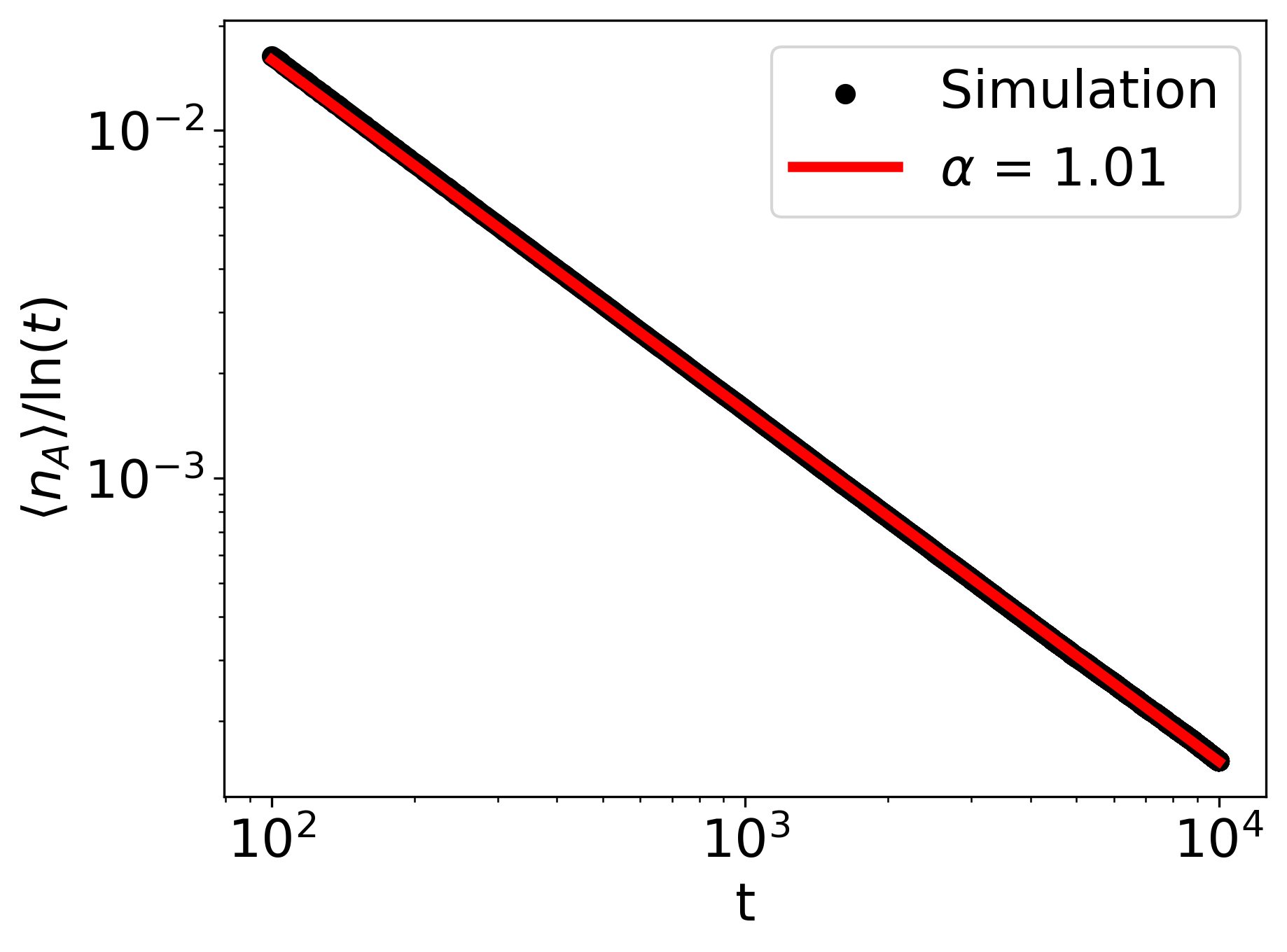}
	\caption{}
    \label{subfig:2d_coagulation_correct}
\end{subfigure}
\hfill
\begin{subfigure}[t]{0.48\textwidth}
    \includegraphics[width=\linewidth]{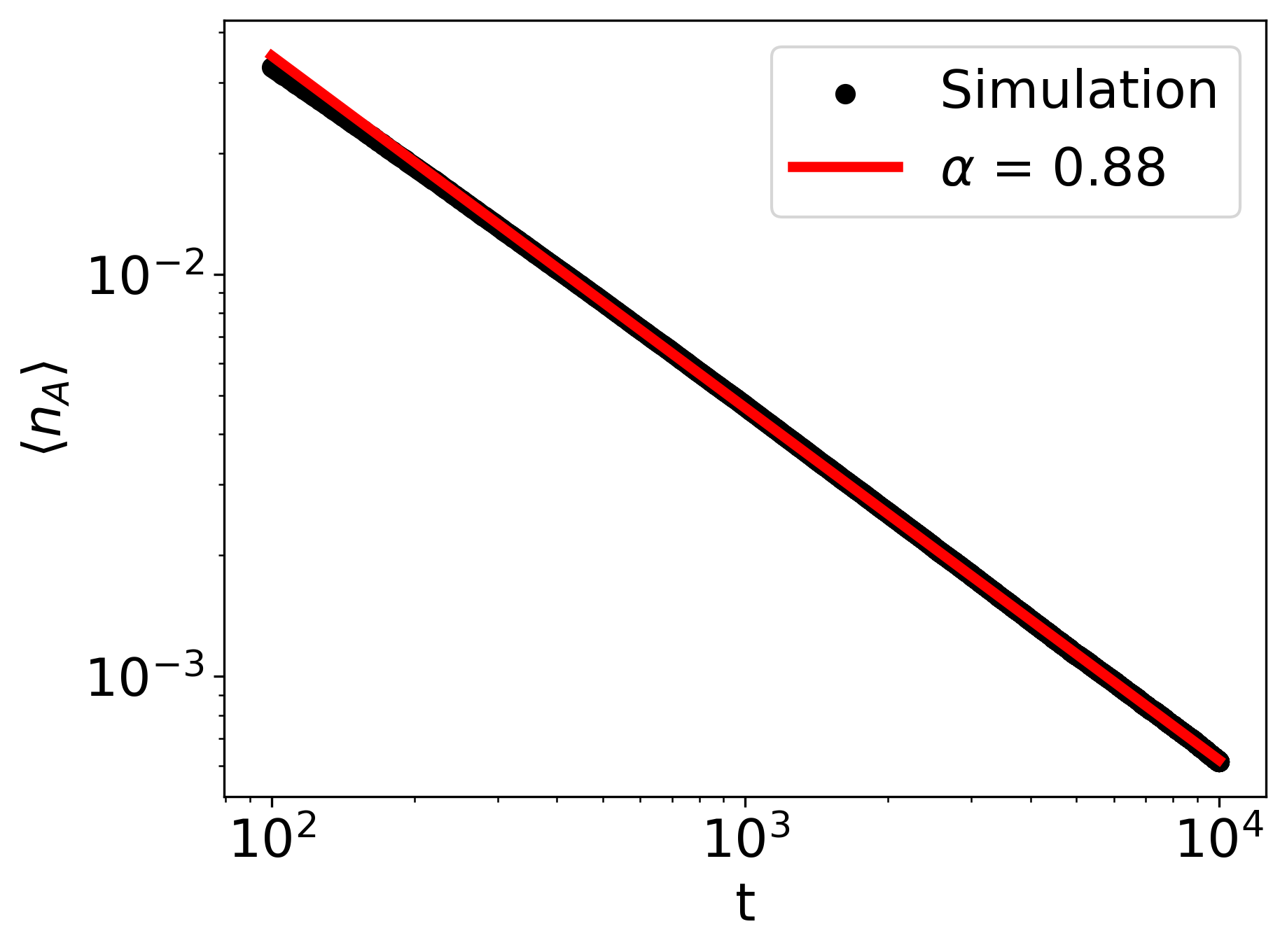}
    \caption{}
	\label{subfig:2d_coagulation_incorrect}
\end{subfigure}
	\caption{Particle density decay for binary coagulation reactions $2 \mathrm{A} \to \mathrm{A}$ with diffusive propagation in  $d=2$ fitted  using the (a) correct scaling with the expected logarithmic correction, and (b)  a pure power-law ansatz. 
    The black line shows the resulting particle density for   simulations on a square lattice with side length $L = 100$, $N_\mathrm{A}(0) = 10^4$ (all sites initially filled), and $\lambda_0 = D_0 = 1$. 
    The data was averaged over $500$ independent runs. 
    The red lines represent fits to the corresponding asymptotic scaling.}
\label{fig:2d_coagulation_density_scaling}
\end{figure}

Monte Carlo data for diffusive binary annihilation processes in  $d=2$ are shown in Fig.~\ref{fig:2d_coagulation_density_scaling}, where the density decay is fitted to a power law taking into account the expected logarithmic correction ($\langle n_A \rangle \sim t^{-1}\ln{t}$)  in Fig.~\ref{subfig:2d_coagulation_correct}, and without the correction in Fig.~\ref{subfig:2d_coagulation_incorrect}. 
Although on first glance both plots seem to lead to a good fit, the exponent $\alpha$ predicted by the incorrect scaling without the logarithm is not consistent with the correct decay exponent $\alpha = 1$ in $d=2$. 
Logarithmic corrections may easily be mistaken as a power law with a slowly varying exponent that differs from the correct one.
This observation underscores the importance of careful data analysis and the fact that prior theoretical input is commonly required to adequately represent simulation (and experimental) results.

\begin{figure}[t]
    \includegraphics[width=0.48\linewidth]{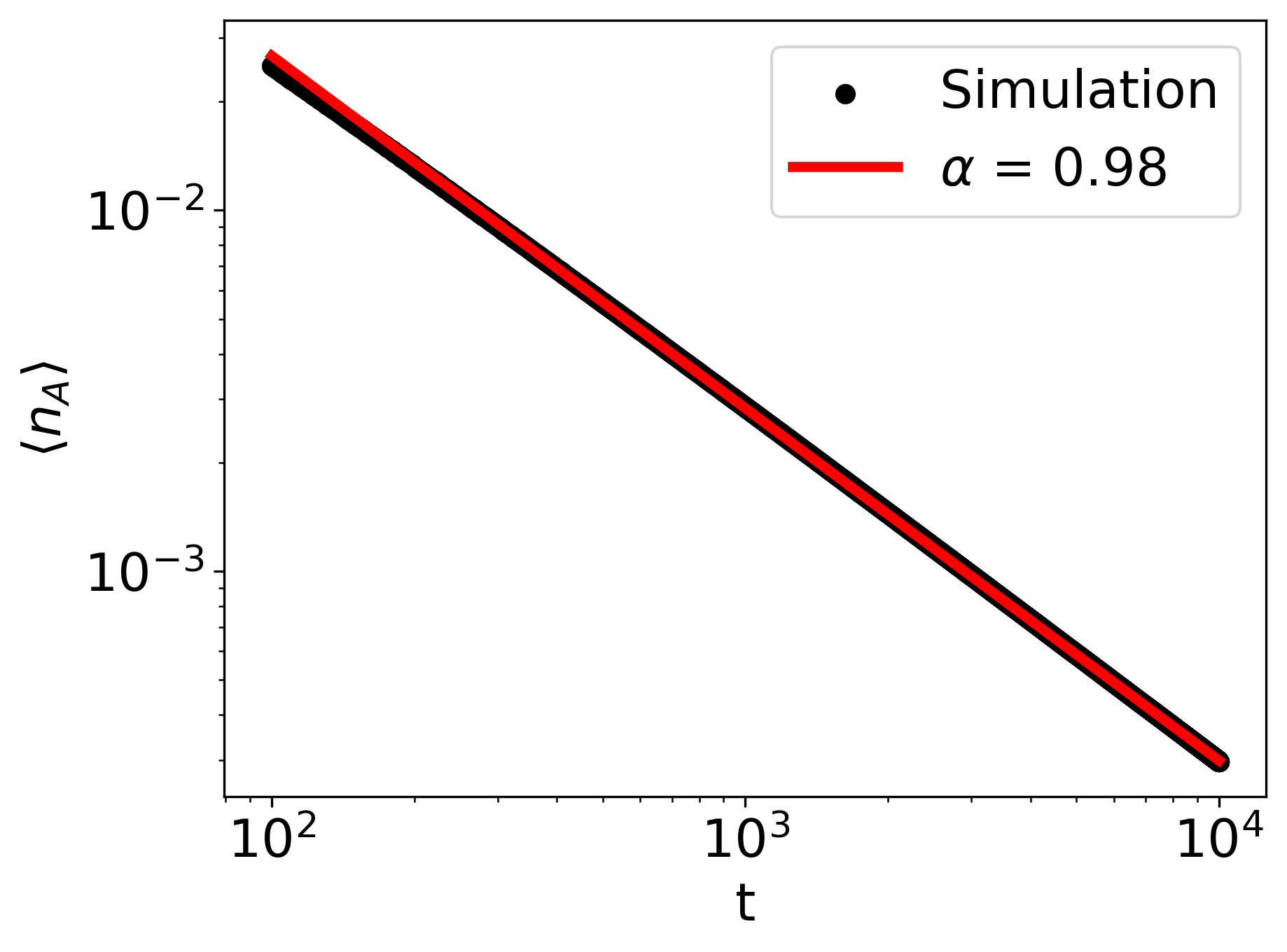}
    \caption{Decay of the particle density for the coagulation reaction $2 \mathrm{A} \to \mathrm{A}$ in $d=3$  fitted to a power law. 
    The black line shows the agent density for Monte Carlo simulations on a cubic lattice with side length $L = 100$, $N_\mathrm{A}(0) = 10^6$ (initially all sites are filled), and $\lambda_0 = D_0 = 1$. 
    The data were averaged over $500$ independent runs. 
    The red line represents a fit to a power law.}
\label{fig:3d_coagulation_density_scaling}
\end{figure}

As the dimension $d$ is increased beyond $d_c = 2$, mean-field reaction-limited scaling takes over. 
For $2 d$ nearest neighbors it takes an ever longer time for a particle to clear a region of radius $\ell_D(t)$, such that the effective reaction rate is dominated only by the annihilation probability.
Figure~\ref{fig:3d_coagulation_density_scaling} shows the density decay in $d=3$, where the data are consistent with the mean-field exponent $\alpha = 1$ (in contrast to the diffusion-limited value $\alpha = 3/2$).
If we fit to the later-time regime, agreement with the expected exponent values can be improved.
We note that extracting local effective scaling exponents by numerically computing the slope in double-logarithmic plots (albeit of  noisy data) helps to assess if the asymptotic long-time regime has been reached.

\subsubsection{Effective rate equation}
\label{ere}
Our discussion suggests that although the mean-field rate equation cannot correctly capture the kinetics of annihilation in systems that are not homogeneous, the density decay is still described by a power law.
In accord with Smoluchowski's analysis in Problem~8, we may assume the existence of an effective rate equation that yields the correct value of $\alpha$. 
This equation can be achieved by ``reverse-engineering'' the rate equation from the known correct time dependence of $\langle n_\mathrm{A}(t) \rangle$ by taking its time derivative.
Instead, we explain the physical interpretation behind this effective rate equation to shed more light on the mean-field approximation.

We first note that the particle number changes only when a coagulation reaction occurs. 
Hence, the time derivative of the density should be proportional to the average number of coagulation events $\langle N_\lambda(t) \rangle$ which occur in a given time interval, that is, the actual (macroscopic) reaction rate, which is a function of $\langle n_\mathrm{A}(t) \rangle$ and hence $t$.
For binary reactions, this reaction rate should be proportional to the reactant density $\langle n_\mathrm{A}(t) \rangle$ at time $t$, as can be directly inferred from the Monte Carlo algorithm described in Sec.~\ref{ssalgo}, where the completion of a Monte Carlo time step is set by the current number of particles in the system.
Thus, we impose $\langle N_\lambda(t) \rangle = \langle \lambda(t) \rangle \langle n_\mathrm{A}(t)\rangle$ with an average per-particle rate of coagulation events $\langle \lambda(t) \rangle$.
Therefore, we can write the  exact rate equation,
\begin{equation}
	\frac{d \langle n_\mathrm{A}(t) \rangle}{dt} = - K  \langle \lambda(t) \rangle \langle n_\mathrm{A}(t) \rangle \ .
\label{eq:effective_coagulation_rate_eq}
\end{equation}
The constant $K$ is determined by microscopic details, but its value is irrelevant, because it can be absorbed into a redefinition of the overall time scale.
Hence we will set $K = 1$.

The effective rate $\langle \lambda(t) \rangle$ can be further decomposed\cite{Swailem2024} by noting that the success of a reaction satisfies two conditions: 
(1) A coagulation occurs according to its probability (the associated randomly generated number needs to be less than $p_\lambda$); and (2) another particle must be present at the site (or within the reaction radius $b$) of the reaction.
Because these events are statistically independent, it follows that $\langle \lambda(t) \rangle = \lambda_0 \langle p_2(t) \rangle$, where $\langle p_2(t) \rangle$ is the average probability for finding two particles on the same site at time $t$.
Although these arguments   are relevant to pair processes, other reaction schemes can be analyzed in a similar manner by introducing an average success rate $R_0$ to encapsulate the first condition.
For a general reaction scheme of the form $\mathrm{A} + \mathrm{B} + \mathrm{C} + \ldots \xrightarrow{R_0} \tilde{\mathrm{A}} + \tilde{\mathrm{B}} +\tilde{\mathrm{C}} + \ldots$, the     effective rate  is $\langle R \rangle = R_0  \langle p(\mathrm{A},\mathrm{B},\mathrm{C},\ldots) \rangle$, where $p(\mathrm{A},\mathrm{B},\mathrm{C}, \ldots)$ is the probability of finding the reactants $\mathrm{A}$, $\mathrm{B}$, $\mathrm{C}, \ldots$ on the same or a neighboring site depending on the Monte Carlo algorithm.

The mean-field approximation may be interpreted as the assumption of a uniform agent density field which may thus be replaced by its mean, allowing us to assume $\langle p_2(t) \rangle \propto \langle n_\mathrm{A}(t) \rangle$.
By choosing the first particle at random, this expression represents the probability that another reactant resides on the same site, assuming a uniform density.
In this approximation, we obtain $d \langle n_\mathrm{A}(t) \rangle / dt \propto - \lambda_0  \langle n_\mathrm{A}(t) \rangle^2$, which is the mean-field rate equation.
However, it becomes clear that this approximation requires the system to be sufficiently homogeneous. 
The existence of spatial correlations would cause $\langle p_2 \rangle$ to deviate from its mean-field value.
The exact calculation of the macroscopic rate $\langle \lambda(t) \rangle$ (or equivalently the average contact probability $\langle p_2 \rangle$) is nontrivial and can be analytically achieved only under special circumstances, usually only in $d=1$.\cite{Tauber2005, Tauber2014}

Instead, we differentiate the density decay law to obtain the following dependencies of $\langle \lambda \rangle$ on the time [see Eq.~(\ref{eq:effective_coagulation_rate_eq})] and on the density:
\begin{subequations}
\begin{align}
	&\langle \lambda(t) \rangle = - \frac{d \ln \langle n_\mathrm{A}(t) \rangle}{d t} \sim t^{-\beta}    &&\beta = 1 \ , \\
	&\langle \lambda(\langle n_\mathrm{A} \rangle) \rangle \sim \langle n_\mathrm{A} \rangle^{\theta}  &&\theta = \beta / \alpha = 1 / \alpha \ .
\end{align}
\label{eq:lambda_scaling_coagulation}
\end{subequations}
Note that the value $\beta = 1$ and hence $\theta = 1 / \alpha$  holds only for  pure power-law scaling and hence requires modification at $d=d_c = 2$.
Determining $\beta$ can be used as a quantitative tool to probe deviations from the asymptotic algebraic scaling regimes.
Mean-field theory yields $\theta = 1$, while diffusion-limited scaling is described by $\theta = 2/d$.
Equation~(\ref{eq:lambda_scaling_coagulation}) relates the rate of events per particle to their density.
By using the Monte Carlo algorithm, we can define a counter that  increases once a coagulation event occurs, and therefore,  simulations may be utilized to directly measure the effective rate $\langle \lambda \rangle$ as a function of time or agent density.
The relation (\ref{eq:effective_coagulation_rate_eq}) is readily verified by numerically computing the density time derivative on the left-hand side as consecutive differences between adjacent discrete data points at each Monte Carlo step and confirming that it equals the product of the mean density and the number of coagulation events per time and per particle.

\begin{figure}[t]
\begin{subfigure}[t]{0.48\textwidth}   
    \includegraphics[width=\linewidth]{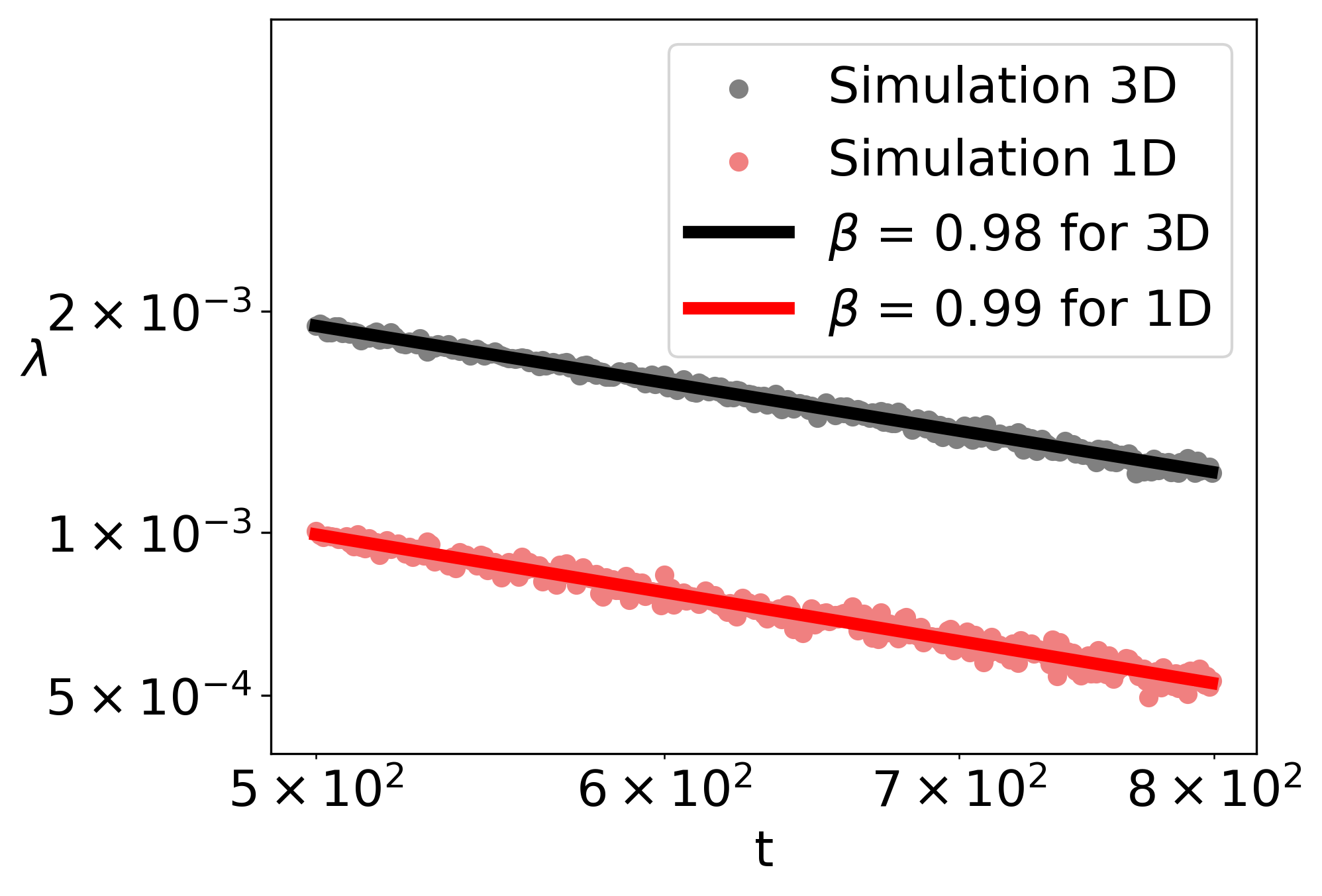}
    \caption{}
    \label{subfig:coagulation_rate_time_scaling}
\end{subfigure}
\hfill
\begin{subfigure}[t]{0.46\textwidth}
    \includegraphics[width=\linewidth]{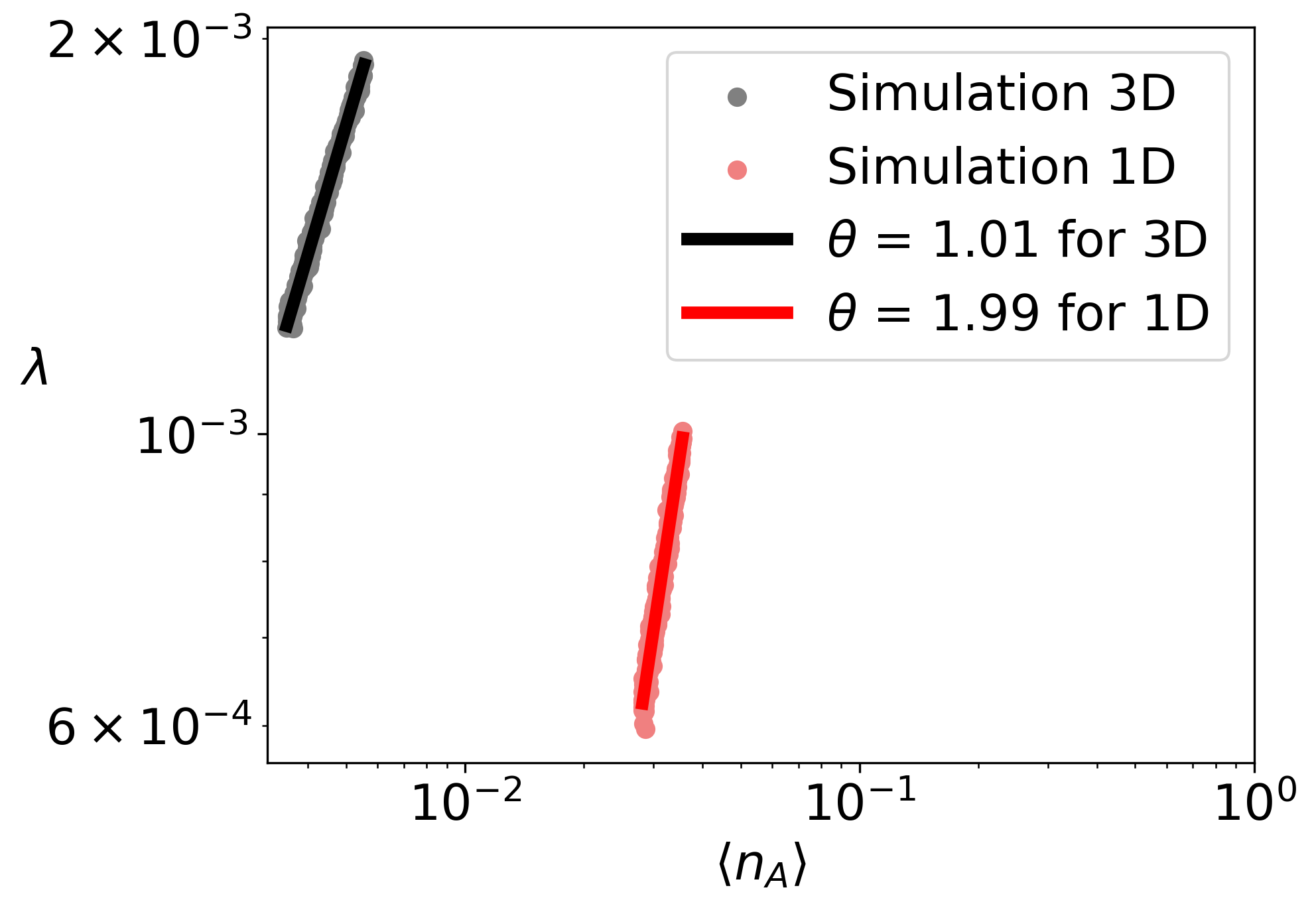}
    \caption{}
    \label{subfig:coagulation_rate_density_scaling}
\end{subfigure}
    \caption{The effective coagulation rate in the binary reaction $2 \mathrm{A} \to \mathrm{A}$ as a function of (a) time and (b) reactant density for $d = 1$ (red) and $d = 3$ (black). 
    The points represent the Monte Carlo data on regular lattices with periodic boundary conditions, and the solid lines represent power law fits. 
    The data on both axes are plotted on $\log_{10}$ scale. 
    }
\label{fig:coagulation_effective_rate}
\end{figure}

The macroscopic reaction rate, which effectively assumes either time or reactant density dependence, is shown in Fig.~\ref{fig:coagulation_effective_rate}.
The value of the exponent $\beta$ does not depend on $d$ and is approximately equal to $1$.
The small deviation from unity is due to correction terms to simple power law scaling whose influence can be reduced by focusing on the data extracted for longer times.
However, at large $t$ the number of particles becomes low and consequently statistical errors induced by the noise are amplified.
The only way to counter this inevitable conundrum and improve the trade-off range is to increase the  system size (number of lattice sites), which is computationally expensive.
The scaling of the effective rate with density exhibits the anticipated behavior with $\theta = 1 / \alpha$, which confirms that for $d = 1$, emerging spatial anti-correlations drastically change the reaction rates that appear in an effective rate equation, and mean-field scaling is recovered above the upper critical dimension $d = 3$, indicating that diffusive propagation efficiently fills the depletion zones and induces a mostly uniform particle distribution.

\begin{itemize}

    \item[] {\bf Problem~9}: Decompose the effective hopping rate $\langle D \rangle = D_0 \, \langle p(\text{reaction\, condition}) \rangle$   in the annihilation/coagulation models with only one particle allowed per site. 
    What would the condition be for the nearest-neighbor hopping reaction to occur? 
    Use your intuition to qualitatively describe the resulting time dependence of the effective hopping rate and do Monte Carlo simulations to check your hypothesis.

\end{itemize}

\subsection{Two-species annihilation}

At first glance, the two-species annihilation reaction $\mathrm{A} + \mathrm{B} \to \emptyset$ would seem to follow that of the single-species annihilation/coagulation processes.
Despite the similarity of these stochastic pair processes, which each require two particles to meet, leading to subsequent annihilation, there is one crucial difference, which is 
the two reactants must be distinct species, because like particles do not affect each other.
This observation suffices to predict a slower density decay for the two-species model, because the reactions are subject to stricter constraints, especially as the particle densities decrease.

We begin our analysis with the coupled mean-field equations for  two species,
\begin{equation}
	\frac{d\langle n_\mathrm{A}(t) \rangle}{dt} = - \lambda_0  \langle n_\mathrm{A}(t) \rangle \langle n_\mathrm{B}(t) \rangle = \frac{d\langle n_\mathrm{B}(t) \rangle}{dt} \ ,
\label{eq:two_species_annihilation_rate_eq}
\end{equation}
where $\lambda_0$ denotes the bare reaction rate.
Although there are two degrees of freedom, there exists a strictly conserved quantity that can be used to eliminate one of them.
The conservation law is apparent because
decreasing the number of $\mathrm{A}$ particles by one also reduces the $\mathrm{B}$ particle count by one.
Therefore, the difference $N_\mathrm{B}(t) - N_\mathrm{A}(t) = C V$ (where $V$ is the system volume) is constant in time, and determined by the initial conditions: $C V = N_\mathrm{B}(0) - N_\mathrm{A}(0)$. 
Without loss of generality, we choose $C > 0$, rendering the $\mathrm{B}$ population the majority species.
On the mean-field level, the conservation law allows us to eliminate $\langle n_\mathrm{B}(t) \rangle = \langle n_\mathrm{A}(t) \rangle + C$ from Eq.~(\ref{eq:two_species_annihilation_rate_eq}):
\begin{equation}
	\frac{d\langle n_\mathrm{A}(t) \rangle}{dt} = - C \lambda_0 \langle n_\mathrm{A}(t) \rangle - \lambda_0 \, \langle n_\mathrm{A}(t) \rangle^2 \ .
\label{eq:two_species_annihilation_single_eq}
\end{equation}
The solution to Eq.~\eqref{eq:two_species_annihilation_single_eq}   can be obtained by separation of variables and integrating over $\langle n_\mathrm{A}(t) \rangle$ using partial fraction decomposition. 
After some algebra, we obtain
\begin{equation}
	\langle n_\mathrm{A}(t) \rangle = \frac{C \, \langle n_\mathrm{A}(0) \rangle e^{- \lambda_0 C t}}{C + \langle n_\mathrm{A}(0) \rangle \left(1 - e^{-\lambda_0 C t}\right)} \ ,
\label{eq:two_species_annihilation_mf_solution}
\end{equation}
which at long times, $\lambda_0 C t \gg 1$, reduces to the exponential density decrease $\langle n_\mathrm{A}(t) \rangle \sim e^{-\lambda_0 C t} \to 0$ and  $\langle n_\mathrm{B}(t) \rangle \to C$ with the characteristic decay time $1 / \lambda_0 C$.

This behavior differs significantly from the single-species annihilation reaction due to the presence of the linear term $- C \lambda_0 \langle n_\mathrm{A}(t) \rangle$ in   Eq.~(\ref{eq:two_species_annihilation_single_eq}), which induces the exponential time dependence, indicating the emergence of the simple decay process $\mathrm{A} \to \emptyset$ with rate $C \lambda_0$. 
At long times, the few surviving minority $\mathrm{A}$ particles are immersed in large inert background clusters of the majority species $\mathrm{B}$.
Consequently, for the isolated $\mathrm{A}$ reactants in an abundant ``sea'' of majority particles, the condition of having to meet a $\mathrm{B}$ particle becomes moot, reducing the pair annihilation reaction effectively to spontaneous ``death'' processes.
This argument relies on the existence of a minority species, and does not apply in the special case when $C = 0$, that is, equal initial densities $n_\mathrm{B}(0) = n_\mathrm{A}(0)$.

\begin{figure}[t]
\begin{subfigure}[t]{0.48\textwidth}
    \includegraphics[width=\linewidth]{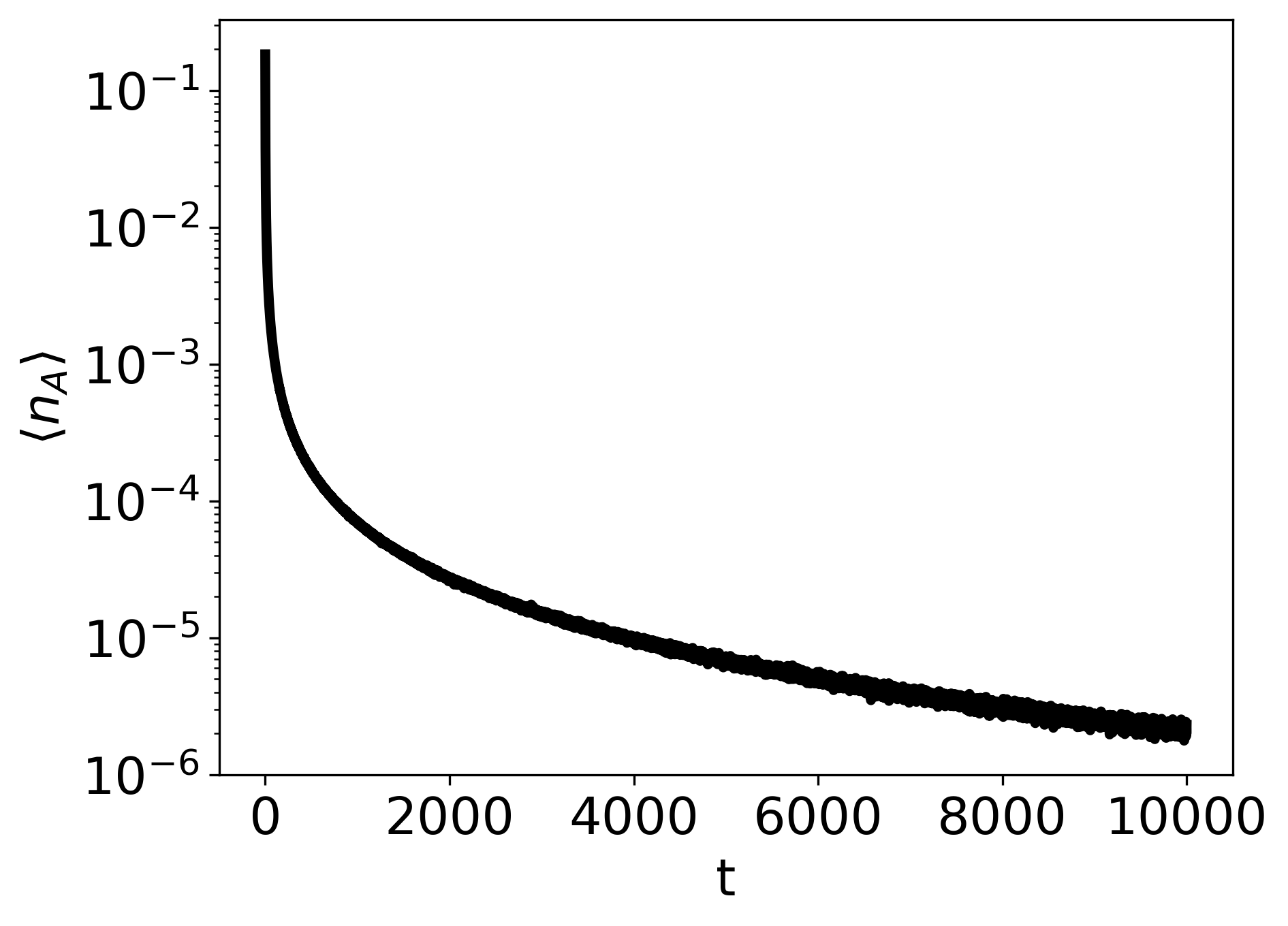}
    \caption{}
    \label{subfig:ann_non_eq_c}
\end{subfigure}
\hfill
\begin{subfigure}[t]{0.48\textwidth}
    \includegraphics[width=\linewidth]{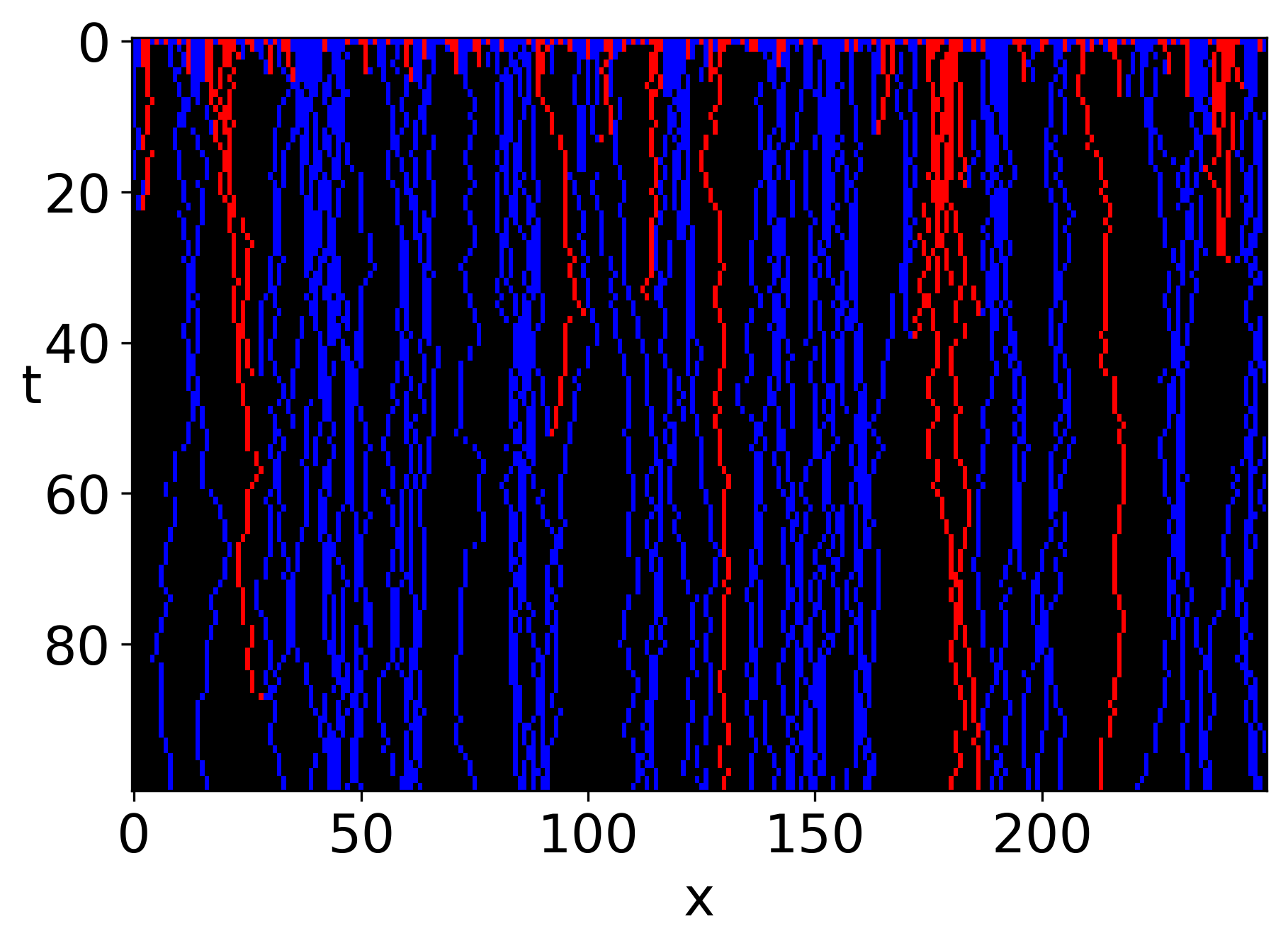}
    \caption{}
    \label{subfig:ann_snapshot}
\end{subfigure}
    \caption{(a) Density  of the minority population for the two-species annihilation reaction $\mathrm{A} + \mathrm{B} \to \emptyset$ on a $d=1$ lattice of length $L = 10^6$ (periodic boundary conditions), with  $N(0) = N_\mathrm{A}(0) + N_\mathrm{B}(0) = L$ (all sites filled), $\lambda_0 = D_0 = 1$, and $C = \big( N_\mathrm{B}(0) - N_\mathrm{A}(0) \big) / L = 0.1$. 
     The density average is taken over $1000$ independent runs. 
    (b) Snapshot of a single simulation run:  
    The horizontal axis represents the position along the lattice, where only the first $250$ lattice sites are depicted. 
    The vertical axis shows the time $t$, running from top to bottom. 
    The world lines of minority particles $\mathrm{A}$ are indicated in red, the majority species $\mathrm{B}$ in blue, and empty sites are displayed black.}
\label{fig:annihilation_non_eq_c}
\end{figure}

We first discuss the case  $C > 0$ and simulate the stochastic two-species pair annihilation reaction $\mathrm{A} + \mathrm{B} \to \emptyset$ on a $d=1$ lattice with periodic boundary conditions to test the validity of the mean-field prediction.
To this end, we plot the logarithm of the minority species density versus time, expecting a straight line for large times if Eq.~(\ref{eq:two_species_annihilation_mf_solution}) holds.
The  data are displayed in Fig.~\ref{fig:annihilation_non_eq_c}, which shows the (a) density decay of the minority species $\mathrm{A}$  and (b) the time dependence  of a single simulation.
The latter demonstrates the development of depletion zones, similar to that of the single-species coagulation reaction in Fig.~\ref{subfig:coagulation_snapshot}, but with significantly fewer empty spaces due to the clustering of majority particles $\mathrm{B}$, which cannot annihilate each other.
It is apparent from the log-linear plot in Fig.~\ref{subfig:ann_non_eq_c} that the density decay does not follow a simple exponential function.
Indeed, anti-correlation effects are expected to modify it to a stretched exponential of the form $\ln \langle n_\mathrm{A}(t) \rangle \sim - (D t)^{d/2}$ in $d < 2$ (with additional logarithmic corrections at $d=d_c = 2$) reflecting the emergence of depletion zones.\cite{Tauber2014}
We also expect pronounced effects due to spatial segregation of the reactive agents into separate inert clusters.

We note that $C = 0$ in the mean-field solution (\ref{eq:two_species_annihilation_mf_solution}) is a singular point in parameter space. 
Hence, we must carefully take the limit $C \to 0$ to obtain
\begin{equation}
    \langle n_\mathrm{A}(t) \rangle = \frac{\langle n_\mathrm{A}(0) \rangle}{1 +\langle n_\mathrm{A}(0) \rangle \lambda_0 t} \ ,
\label{eq:mfpairan}    
\end{equation}
which is just the mean-field rate  solution for the single-species binary annihilation reaction.
The limits $t \to \infty$ and $C \to 0$ do not commute.

\begin{figure}[t]
\begin{center}
\includegraphics[width=0.48\textwidth]{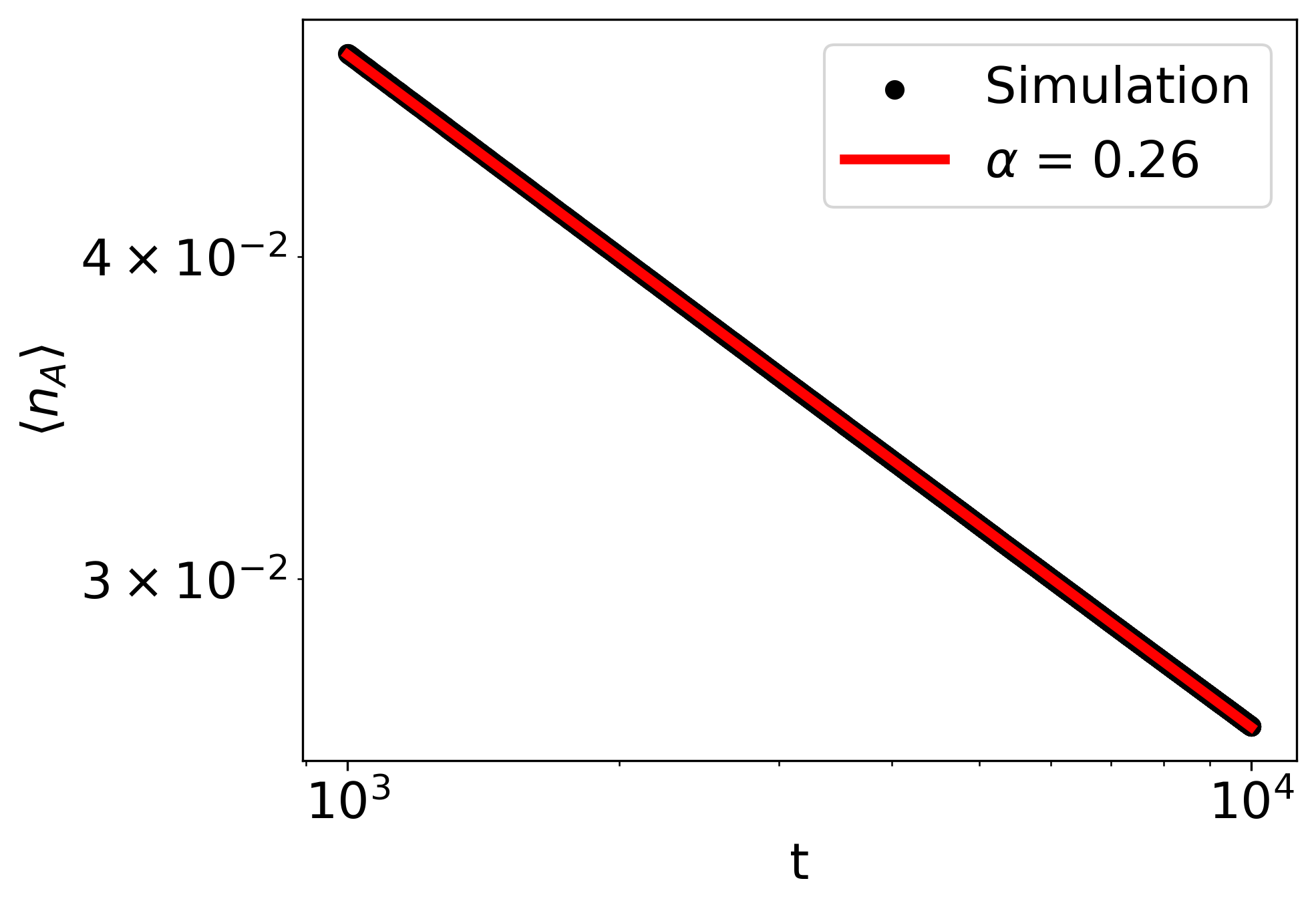}
\end{center}
    \caption{Particle density decay for the two-species pair annihilation reaction $\mathrm{A} + \mathrm{B} \to \emptyset$ in $d=1$  with equal $\mathrm{A}$ and $\mathrm{B}$ densities ($C = 0$) fitted to a power law. 
     The black line shows the particle density  from   simulations in $d=1$ with $L = 10^5$ sites, $N(0) = L$, $\lambda_0 = D_0 = 1$. 
    The data were averaged over $500$ independent runs. 
    The red plot is a fit of the data to a simple power law with decay exponent $\alpha$.}
\label{fig:ann_c_0}
\end{figure}

From the mean-field perspective, the single- and two-species binary annihilation models become equivalent when $C=0$, and there is no difference between the $\mathrm{A}$ and $\mathrm{B}$ particle densities.
In this case the mean density field created by either species becomes uniform and equal to the spatially averaged particle density.
Correspondingly, the mean-field assumption ignores the fact that within either $\mathrm{A}$- or $\mathrm{B}$-rich regions, no reactions occur, so that the system's annihilation dynamics is confined to the interfaces separating these clusters.
By exploiting the conservation law for the particle number difference, we conclude that the corresponding local field $c(\vec{x},t) = n_\mathrm{B}(\vec{x},t) - n_\mathrm{A}(\vec{x},t)$ must   satisfy a simple diffusion equation, without any contributions from the pair reactions.
A straightforward analysis based on  Gaussian statistics for diffusive processes yields that any local species density excesses behaves as $\overline{|c(\vec{x},t)|} \sim (D t)^{-d/4}$ in $d$ dimensions,~\cite{Tauber2005,Tauber2014} setting the limiting scale for binary annihilations for $d < d_s = 4$, the boundary dimension for species segregation to be maintained in the presence of diffusive mixing.
The  predicted density decay $\langle n_\mathrm{A/B}(t) \rangle \sim (D t)^{-d/4}$ is even slower than the depletion-induced single-species power law and determines the asymptotic kinetics for $d < d_s = 4$. 
At intermediate times, we expect more generally that $\langle n_\mathrm{A/B}(t) \rangle = A  t^{-\alpha} + A' t^{-\alpha'}$, with the segregation exponent $\alpha = d/4$ and the depletion exponent $\alpha' = \textrm{min}(1,d/2)$.
Figure~\ref{fig:ann_c_0} displays Monte Carlo data for the $\mathrm{A} + \mathrm{B} \to \emptyset$ reaction with diffusive particle spreading in $d=1$.
Fitting to a single algebraic decay yields the decay exponent $\alpha \approx 0.26$.
The small upward deviation from the expected segregation value $\alpha = 1/4$ likely originates from a second contribution with $\alpha' = 1/2$.
For $d > 1$. simulations on very large lattices  are required to reach the true long-time decay regimes and acquire data with sufficient statistical significance over the inevitable small numbers of particles.\cite{Leyvraz1992, Cornell1992}

\subsection{Spontaneous structure formation in predator-prey models}

\begin{figure}[t]
\begin{center}
    \includegraphics[width=0.5\textwidth]{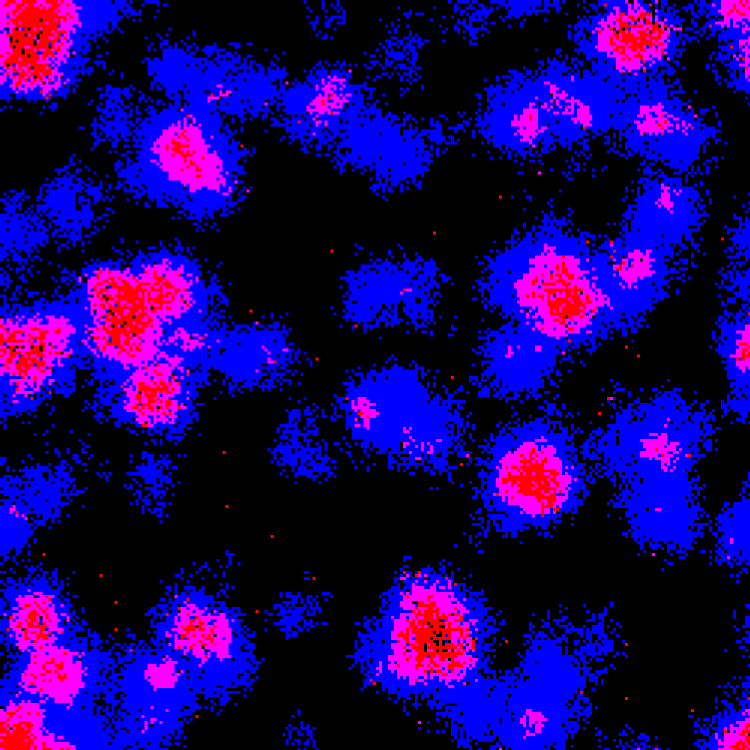}
\end{center}
    \caption{Simulation snapshot of a standard Lotka--Volterra predator-prey competition model on a $250 \times 250$ square lattice  with rates $\sigma = \mu = \lambda = 0.1$ and initial densities $n_\mathrm{A} = 0.1$ and $n_\mathrm{B} = 1$. 
    Patches of prey  $\mathrm{B}$ (blue) are invaded by predators $\mathrm{A}$ (red), with emerging activity fronts clearly visible. 
    Multiple site occupations are permitted. Lattice sites containing members of both species are colored pink.}
\label{fig:lv-activity-fronts-patches}
\end{figure}

Adversarial multi-species models exhibit a wide range of spontaneously emerging spatial patterns. 
A prime example is the activity fronts that form when particles in the Lotka--Volterra predator-prey model (see Problem~4) are allowed to move on a $d=2$ or $d=3$ lattice. 
A spreading activity front appears when comparatively few predator particles $\mathrm{A}$ enter an area abundant with prey $\mathrm{B}$. 
On this front, $\mathrm{A}$ particles can reproduce rapidly, inducing local population expansion. 
Prey $\mathrm{B}$ are consumed during this process, leaving a depletion region of low abundance in which $\mathrm{A}$ cannot survive for long. 
Hence, the bulk of surviving predators necessarily moves toward areas of high prey abundance, leaving behind empty spaces in the lattice where over time species $\mathrm{B}$ can replenish, thus enabling another activity front to pass through.
This intriguing nontrivial persistent wave front dynamics can be readily observed in a snapshot of the Lotka--Volterra model, as displayed in Fig.~\ref{fig:lv-activity-fronts-patches}. 
In finite systems, the activity waves continue to spread and interact with each other, inducing persistent (but random) population oscillations.\cite{Mobilia2007, Dobramysl2018, Tauber2024}
This behavior is an example of the emergence of noise-stabilized spatio-temporal features in nonequilibrium systems.
Based on our discussion of species segregation for $\mathrm{A} + \mathrm{B} \to \ldots$ pair reactions with distinct species, we expect prominent activity fronts to dominate the system for $d < d_s = 4$, which is  the case.\cite{Mobilia2006} 

The activity front speed and size are linked to the time and length scales of the system. 
It is  interesting to study how quickly these spreading wave fronts propagate and to determine the dependence of the system's mesoscopic scales on the model parameters.
Unless the model is carefully tuned to allow for activity fronts to be sufficiently well-defined and extended, it is difficult to identify and track them over appreciable time intervals as new fronts continually appear and merge, rendering them potentially very short-lived, as is apparent in Fig.~\ref{fig:lv-activity-fronts-patches}. 
In the following, we simplify the task of measuring the spreading interaction front of predators $\mathrm{A}$ invading into a sea of non-reproductive and stationary prey species $\mathrm{B}$. 
To this end, we completely fill the lattice with prey and set their reproduction and hopping rates to zero. 
We then put a small number of $\mathrm{A}$ particles in the ``center'' of the lattice and observe the ensuing activity front spread into the occupied $\mathrm{B}$ region. 

\begin{itemize}

    \item[] {\bf Problem 10}: The algorithm for the spatial Lotka--Volterra model is implemented as an interactive, browser-based Python Notebook.\cite{github3} 
    Familiarize yourself with the notebook and run the example without modifying it.
    
    \item[] {\bf Problem 11}: Change the predator death rate in Problem~10 to $\mu = 0.5$ and observe how the front speed changes as a function of the interaction rate $\lambda$.
    
    \item[] {\bf Problem 12}: Modify the notebook to determine not only the activity front speed, but also the spatial width of the predator wave.
    
    \item[] {\bf Problem 13}: Devise a method to determine the characteristic distance between individuals of species $\mathrm{A}$ over time using the autocorrelation function. 
    (Hint: this method likely requires averaging the autocorrelation function over many runs. Thus, it may be beneficial to set up a local Python environment rather than running the notebook in a browser.)

\end{itemize}

\subsection{Evolutionary ecological dynamics}

The paradigmatic Lotka--Volterra model for predator-prey coexistence is comparatively easy to understand and allows for interesting biologically relevant extensions, such as the implementation of spatial variability,\cite{Dobramysl2008} periodically varying external resource limitations,\cite{Swailem2023} and other extensions.\cite{Dobramysl2018}
The co-evolution of survival traits of competing predator and prey species is another intriguing example.\cite{Dobramysl2013a, Dobramysl2013b}
We briefly discuss the treatment of the mutual adaptation of two interacting populations by extending the basic stochastic Lotka--Volterra model and its analysis using agent-based simulations.

Predators $\mathrm{A}$ are assigned a trait value $z_{\mathrm{A},i}$ between zero and one, which loosely captures how efficient they are at hunting prey. 
The index $i$ identifies the predator.
Prey $\mathrm{B}$ are assigned a similar trait value $z_{\mathrm{B},j}$, which describes their proficiency at evading predators and $j$ identifies a specific prey. 
When $\mathrm{A},i$ interacts with $\mathrm{B},j$, their competitive trait values are combined somewhat arbitrarily by their arithmetic mean to form the probability of a successful predation interaction: $\lambda_{ij} = (z_{\mathrm{A},i} + z_{\mathrm{B},j}) / 2$. 
Therefore, it is beneficial for predators to have $z_{\mathrm{A},i}$ close to one (giving them a higher chance to consume their prey and reproduce), and for prey to have $z_{\mathrm{B},j}$ close to zero to avoid being devoured.

To close the cycle of adaptation, we need to define a way to assign trait values to agents beyond the initial conditions. 
When offspring of either species are created (for species $\mathrm{A}$ by a predation interaction, and for prey $\mathrm{B}$ by spontaneous birth, there is always a parent associated with this newly emerging agent. 
We can use the parent's trait value $z_\mathrm{P}$ as a reference and choose its offspring's trait value $z_\mathrm{O}$ to be close to this reference number, thereby allowing individuals to pass on their efficacy to their progeny. 
We choose $z_\mathrm{O}$ from a Gaussian distribution with mean $z_\mathrm{P}$ and width $\nu_\mathrm{A/B}$, truncated to the interval $[0,1]$. 
Here, $\nu_{A}$ and $\nu_{B}$ measure the strength of mutation variabilities, or effective mutation rates for species $\mathrm{A}$ and $\mathrm{B}$, respectively. 
These parameters determine how quickly either population changes its overall trait distribution by the combination of competitive and reproductive stochastic processes, and the amount of variability in the trait values at the resulting stationary state. 
This choice of trait inheritance is arbitrary (see Problem~16).
This still rather simple model shows very interesting dynamics in the distribution of traits in both species' populations. 
If the parameters are chosen such that coexistence is possible, predators and prey co-evolve until their trait distributions reach a steady state. 
During this evolutionary process, both species' abundances change with the adaptation, allowing slightly enhanced population numbers compared to the corresponding adaptation-free system.\cite{Dobramysl2013b}

\begin{itemize}

    \item[] {\bf Problem 14}: Open the Python Notebook\cite{github4} and note how the rules are implemented. 
    Then run the simulations and observe the results. 

    \item[] {\bf Problem 15}: Vary the mutation rates $\nu_\mathrm{A}$ and $\nu_\mathrm{B}$ and determine how these alterations influence the rate of adaptation and the ensuing specialization of both species, that is, how much trait values vary in the population in the quasi-stationary state.

    \item[] {\bf Problem 16}: Implement another way to assign a trait value by replacing the truncated Gaussian distribution with a different distribution. 
    Observe how the variation affects the adaptation dynamics.

\end{itemize}

\subsection{Population extinctions as non-equilibrium phase transitions}

The complexity of species interaction networks subject to continually changing and spatially inhomogeneous environmental conditions render analyzing the evolutionary dynamics of natural populations a challenging task.\cite{May1973, Maynard-Smith1974, Maynard-Smith1982, Hofbauer1998, Murray2002}
One of the first attempts to characterize large ecological systems dates more than fifty years ago, when the stability of a system with random interspecies interactions was investigated by random matrix theory.\cite{May1972} 
Since then, many studies have addressed the evolution of competing and coexisting populations, typically by considering sets of coupled differential rate equations. 
These studies include the  Lotka--Volterra model for hierarchical predator-prey type interactions and variants that describe cyclical dominance of three competing species, namely, the Lotka--Volterra or rock-paper-scissors and the May--Leonard models.\cite{Maynard-Smith1982}
These mean-field models cannot account for population fluctuations, which are usually non-negligible in ecosystems because typical population sizes are much smaller than the sizes of condensed-matter systems.
In addition, reaction-induced correlations may generate emerging spatio-temporal structures. 
The deterministic rate equations also fail near absorbing states, and hence cannot adequately characterize the key properties of a population near extinction or fixation.\cite{Hinrichsen2000, Henkel2008, Tauber2014, Lindenberg2020} 

Absorbing states in population dynamics are reached by the extinction of one species, because once the number of a species has decayed to zero, this population cannot recover. 
In a  two-species ecology, the other population then reaches fixation.
The sharp transition between the active and extinction states is characterized by singularities in the system's macroscopic observables which closely resemble the phenomenology at critical points in thermal equilibrium.
At an extinction critical point, the gain and loss of individuals in the vulnerable population are exactly balanced. 
Because the population cannot escape the absorbing state once it has been reached, the detailed balance condition with any other active state is manifestly violated, and the system displays an out-of-equilibrium phase transition.
To investigate the properties of active-to-absorbing state transitions, we can determine how the ``active'' species density $\langle n_a(t,\tau) \rangle$ (which serves as the order parameter for the transition) grows on the active side, how it decays with time at criticality,  and how the characteristic correlation lengths $\xi(\tau)$ and relaxation times $t_c(\tau)$ diverge in terms of the control parameter $\tau$ as the extinction threshold is approached:\cite{Hinrichsen2000, Odor2004, Tauber2005, Henkel2008, Tauber2014} We write
\begin{subequations}
\begin{align}
    &\langle n_a(t \to \infty, \tau) \rangle \sim \tau^\beta &&
    \langle n_a(t, \tau = 0) \rangle \sim t^{- \alpha} \ , \\ 
    & \xi(\tau) \sim |\tau|^{-\nu} &&
    t_c(\tau) \sim \xi(\tau)^z \sim |\tau|^{- z \nu} \ ,
\label{stcrex}    
\end{align}
\end{subequations}
which defines the critical exponents $\beta$, $\alpha$, $\nu$, and $z$.
Only three of these exponents are independent and they are related by the scaling relation $\alpha = \beta / z \nu$.\cite{Hinrichsen2000, Odor2004, Tauber2005, Henkel2008, Tauber2014}

Determining these quasi-stationary exponents may be impeded by finite-size and accompanying finite run-time limitations.
An efficient approach is to resort to seed simulations and  set the system close to criticality ($\tau \approx 0$) and follow the  evolution of a single small cluster of active sites with time  through the initial increase of the mean number of active sites $\langle n_a(t) \rangle$, the decay of the active cluster survival probability $P_s(t)$ (requiring at least one active individual at time $t$), and the growth of the correlation length:
\begin{equation}
    \langle n_a(t) \rangle \sim t^\theta \ , \quad 
    P_s(t) \sim t^{- \delta} \quad 
    \xi(t) \sim t^{1/z} \ ,
\label{sdcrex}    
\end{equation}
with new critical exponents $\theta$, $\delta$, and the dynamic critical exponent $z$ defined in Eq.~(\ref{stcrex}).
Active-to-absorbing state phase transitions are captured by the directed percolation universality class.\cite{Hinrichsen2000, Odor2004, Janssen2005, Henkel2008, Henkel2011, Tauber2014}
This universality assertion holds even for multi-species systems and applies to the predator extinction threshold which emerges in the Lotka--Volterra model with finite prey carrying capacity.\cite{Mobilia2007, Tauber2012, Chen2016, Dobramysl2018, Tauber2024}
For critical directed percolation, the scaling relations $\delta = \alpha$ and $\theta + 2 \alpha = d / z$ connect the seed and stationary scaling exponents.

\subsection{Epidemic spreading}

The Susceptible-Infectious-Recovered (SIR) model is a paradigmatic but simplified model for epidemic spreading with recovery (or death),\cite{Murray2002} and serves as a modeling work horse in epidemiology, where many variants have been investigated, especially during the Covid-19 pandemic.
The  population in the SIR model is divided into three species or compartments: susceptible $\mathrm{S}$, infected and infectious $\mathrm{I}$, and recovered and immune to infection $\mathrm{R}$. 
The disease spreading dynamics is represented by two irreversible state transitions, the infection reaction $\mathrm{S} + \mathrm{I} \to \mathrm{I} + \mathrm{I}$, which occurs with rate $r$, and the recovery process $\mathrm{I} \to \mathrm{R}$ with rate $\gamma$. 
Note that the binary infective process is the predation reaction in the Lotka--Volterra model for predator-prey competition. 
However, in SIR dynamics, agents may change their species identity, but the total population $N_\mathrm{S} + N_\mathrm{I} + N_\mathrm{R} = N$ is conserved.

If we apply standard mass action-type factorization to the two-body correlations in the infection process, we recover the mean-field  differential rate equations for the three species,
\begin{subequations}
\label{sirdyn}
\begin{align}
    \frac{d \langle n_\mathrm{S}(t) \rangle}{dt} &= - \beta  \frac{\langle n_\mathrm{I}(t) \rangle \, \langle n_\mathrm{S}(t) \rangle}{N} \ , \\
    \frac{d \langle n_\mathrm{I}(t) \rangle}{dt} &= \beta  \frac{\langle n_\mathrm{I}(t) \rangle \, \langle n_\mathrm{S}(t) \rangle}{N} - \gamma \langle n_\mathrm{I}(t) \rangle \ , \\
    \frac{d \langle n_\mathrm{R}(t) \rangle}{dt} &= \gamma  \langle n_\mathrm{I}(t) \rangle \ ,
\end{align}
\end{subequations}
where $\beta \sim r N$ is the continuum mean-field infection rate, scaled by the total population. 
Typical initial conditions are $\langle n_\mathrm{R}(0) \rangle = 0$ and $\langle n_\mathrm{I}(0) \rangle \ll \langle n_\mathrm{S}(0) \rangle \approx N$. 
The basic reproduction number  $\mathcal{R}_0 = \beta / \gamma$  characterizes the solutions of Eq.~\eqref{sirdyn} and is an important epidemiological parameter that  predicts an epidemic outbreak if $(d \langle n_\mathrm{I} \rangle / dt)|_{t=0} > 0$, that is, for $\mathcal{R}_0 > \mathrm{N} / \langle n_\mathrm{S}(0) \rangle \approx 1$. 
The disease kinetics is governed by the behavior of the susceptible and infectious species, because $\langle n_\mathrm{R}(t) \rangle = \langle n_\mathrm{R}(0) \rangle + \gamma \int_0^t \langle n_\mathrm{I}(t') \rangle \, dt'$.

Although the SIR mean-field rate equations can accurately capture the course of an epidemic near its peak and for large and well-connected populations,\cite{Murray2002} they neglect two-point (and higher) correlations, and consequently cannot adequately account for spatio-temporal fluctuations. 
The latter stem from environmental variability and the intrinsic stochasticity of the reaction processes, and become important quantitatively at the outset and at the late stage of the epidemic.
Stochastic fluctuations may drive the system into the absorbing epidemic extinction state during the early stages of the infection outbreak and crucially affect the scaling properties at the epidemic threshold.\cite{Tauber2014, Tauber2020, Mukhamadiarov2021} 

We focus on the stochastic implementation of the SIR reactions $\mathrm{S} + \mathrm{I} \xrightarrow{r} \mathrm{I} + \mathrm{I}$ and $\mathrm{I} \xrightarrow{a} \mathrm{R}$ by simulating them on a $d$-dimensional lattice. 
We employ Monte Carlo simulations with random sequential updates, starting with a single infectious $\mathrm{I}$ seed at the lattice center, and the rest of the population set to the susceptible $\mathrm{S}$ state.
We proceed by randomly selecting lattice sites and attempting to perform the SIR reactions. 
To complete one Monte Carlo step, we repeat this procedure of random site selection and reaction attempts $N=L^d$ times.

\begin{figure}[t]
   \includegraphics[width=\textwidth]{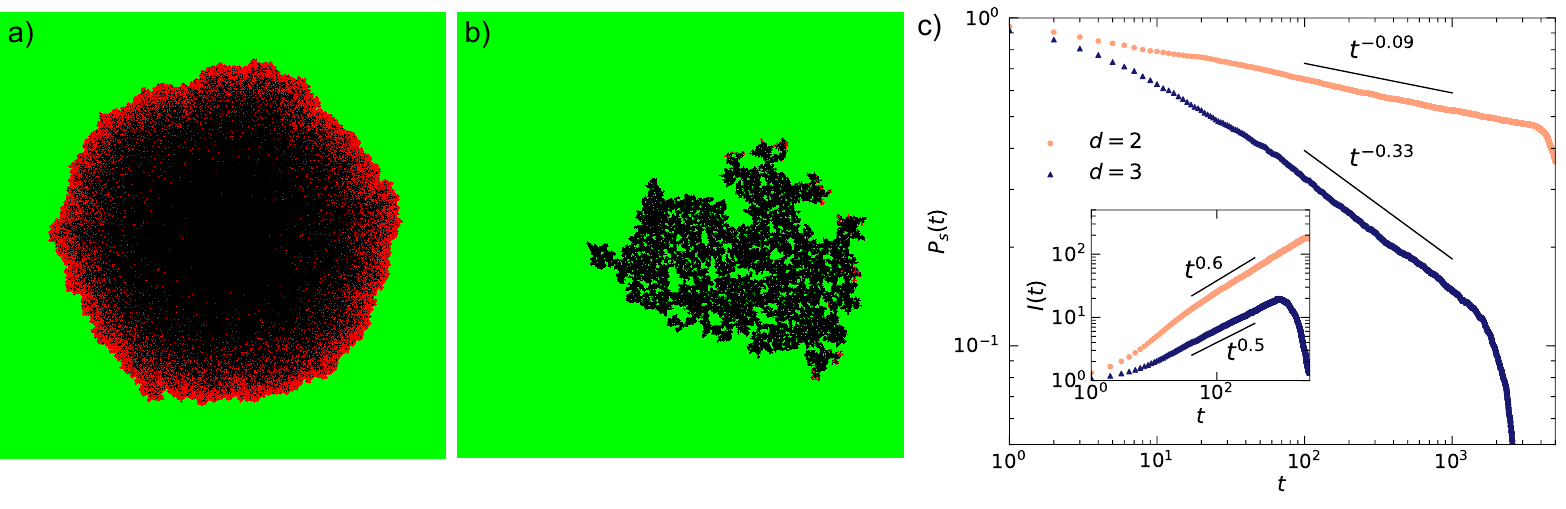}
    \caption{Monte Carlo  snapshots of  Susceptible-Infectious-Recovered (SIR) epidemic dynamics  on a square lattice ($L = 500$) with color coding $\mathrm{S}$: green, $\mathrm{I}$: red, $\mathrm{R}$: black.
    (a) Front expansion in the epidemic spreading phase ($r = 1.0$, $a = 0.1$). 
    (b) Emerging percolation cluster near the critical epidemic threshold ($r = 0.105$, $a = 0.1$).
    (c) Evolution of the disease survival probability $P_s(t)$ and the mean number of infected individuals $\langle n_\mathrm{I}(t) \rangle$ measured at the critical point on ($d = 2$, $L = 500$) and cubic ($d = 3$, $L = 50$) lattices. 
    Both quantities display power laws with critical dynamic isotropic percolation scaling exponents until the finite-size cutoff is reached at large times $t$; the data were averaged over $3,000$ independent runs. 
Reproduced with permission from Ref.~\onlinecite{Mukhamadiarov2022b}.}
\label{fig:SIR}
\end{figure}

For different values of the infection and recovery rates, the disease either spreads throughout the system or remains contained within localized infected clusters.
These distinct macroscopic states are separated at the disease extinction threshold by a continuous non-equilibrium active-to-absorbing phase transition. 
As shown in Fig.~\ref{fig:SIR}(a), the dynamics in the spreading or active phase, as seeded by a few neighboring infectious sites, takes the form of a growth front emanating from the central seed and expanding into the entire domain originally populated by susceptible individuals. 
In contrast, in the non-spreading or inactive phase, the system quickly reaches the absorbing state with all infectious individuals recovering, thus leaving the bulk of the susceptible population untouched by the disease. 
At the extinction critical point, the disease cluster spreads as displayed in Fig.~\ref{fig:SIR}(b), and the population densities, the disease survival probability $P_s(t)$, and the mean-square displacement of the spreading disease from its origin $\langle \mathcal{R}^2(t) \rangle$ exhibit power law scaling similar to Eq.~(\ref{sdcrex}):
\begin{subequations}
\begin{align}
    N - \langle n_\mathrm{S}(t) \rangle &\sim \langle n_\mathrm{R}(t) \rangle \sim t^{\theta_\text{R}} && \langle n_\mathrm{I}(t) \rangle \sim t^{\theta} \ , \\
    P_s(t) &\sim t^{-\delta} x &&\langle \mathcal{R}^2(t) \rangle \sim \xi(t)^2 \sim t^{2/z} \ ,
\end{align}
\label{crtexp}
\end{subequations}
with scaling exponents $\delta$, $\theta$, and $\theta_\text{R}$, and the associated dynamical critical exponent $z$.
For the SIR non-equilibrium phase transition, these exponents are determined by the dynamic isotropic percolation universality class.\cite{Janssen2005, Henkel2008, Tome2010, Tauber2014}.
Examples of numerically extracting power laws for the SIR model in $d = 2$ and $d = 3$ are shown in Fig.~\ref{fig:SIR}(c).\cite{Mukhamadiarov2022b}

In contemporary human populations, epidemic spreading in a confined spatial environment is more realistically described by a network with the topology and connectivity that represents social contact 
interactions.\cite{Pastor2015}
The study of infectious kinetics on a variety of network structures, including random graphs, scale-free, and small-world networks, has helped epidemiologists better understand the dynamic properties of contagious disease propagation.\cite{Pastor2015}

Due to the Covid-19 pandemic, a wide range of compartmental epidemic models have been investigated, each capturing a particular relevant aspect of the disease spreading dynamics. 
For example, if the recovery process has a memory and recovered individuals lose their immunity over time, we can use the Susceptible-Infections-Recovered-Susceptible (SIRS) model,\cite{Anderson1992, Murray2002, Keeling2011} or in the extreme case of very short-lived immunity, the SIS model, whose critical dynamics falls into the directed percolation universality class.\cite{Henkel2008}
If incubation times are relevant, we can introduce an exposed state $\mathrm{E}$ that is not yet infectious, and consider Susceptible-Exposed-Infected-Recovered (SEIR) models.\cite{Anderson1992, Murray2002, Keeling2011} 
Similarly, vaccination effects can be captured by an additional vaccinated state $\mathrm{V}$ and a $S \to V$ transmutation reaction (SIRV model).\cite{Keeling2011}

\section{ Tips and Caveats}
\label{tipcav}

We have provided an   overview of the importance of stochasticity in different contexts, ranging from simple reaction-diffusion systems to more complex multi-species models, and discussed how to perform stochastic simulations of such models.
These simulations need to be done with care, because the data  is inevitably noisy, and it is often difficult to distinguish between a genuine signal or an artifact of the simulation.
We now discuss some tips and caveats that we have learned 
in the hope of helping readers to avoid some common pitfalls.

\subsection {Statistically significant data and error estimates}

Because the phenomena we have considered are subject to inherent randomness, single stochastic  trajectories tend to be meaningless.
We might be tempted to  compute the full probability distribution of the state space of the system, which encapsulates any quantity of interest. 
Such a determination is rarely feasible due to the huge number of possible paths through configuration space which the system can take.
Rather, the usefulness of simulations resides in our ability to compute configurational averages and probabilities of specific events (for example, extinction or fixation).
Any quantity of interest should be computed for sufficiently many individual trajectories to ensure that the results are statistically significant and reach the desired accuracy.
That is, if another researcher were to perform simulations of the same model (or ideally, do a corresponding laboratory experiment), the obtained averaged observables should be consistent within their error bars.

In contrast to systematic errors or deviations between models that originate from the algorithms, the statistical significance of simulation data may always be improved by doing more realizations. 
Because this improvement comes with significant additional computational cost, we provide some tips to guide the determination of the minimum required  number of realizations to yield a statistically significant result for the quantity of interest, and  offer some caveats that apply specifically to systems with  absorbing states.

\begin{figure}[t]
    \begin{subfigure}[t]{0.47\textwidth}	      
    \includegraphics[width=\linewidth]{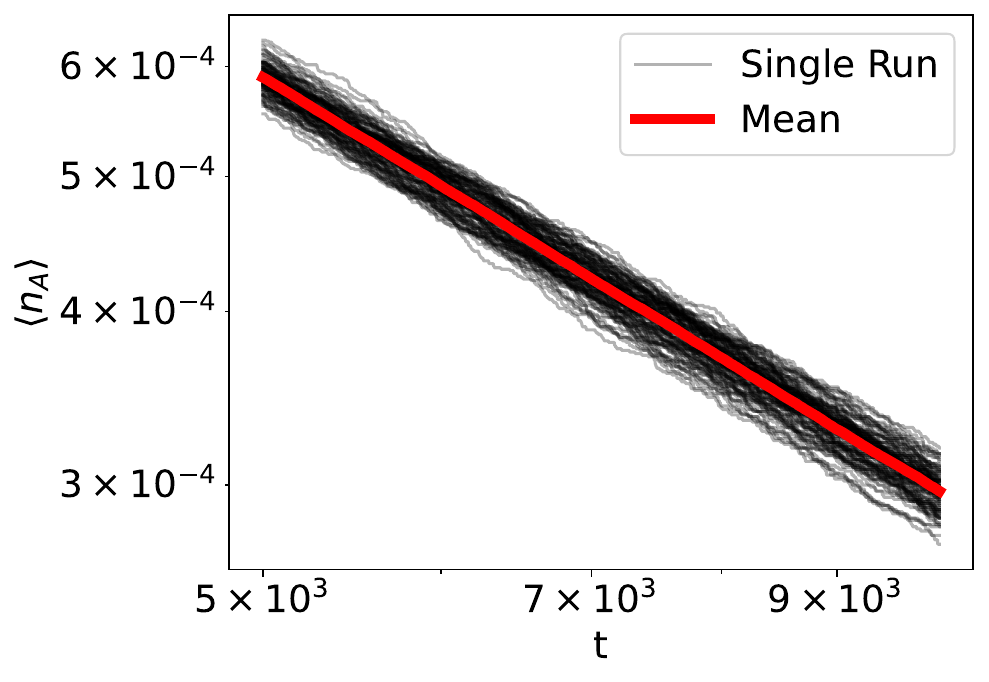}
    \caption{}
    \label{subfig:cog_3d_runs_with_mean}
    \end{subfigure}
    \hfill
    \begin{subfigure}[t]{0.48\textwidth}
    \includegraphics[width=\linewidth]{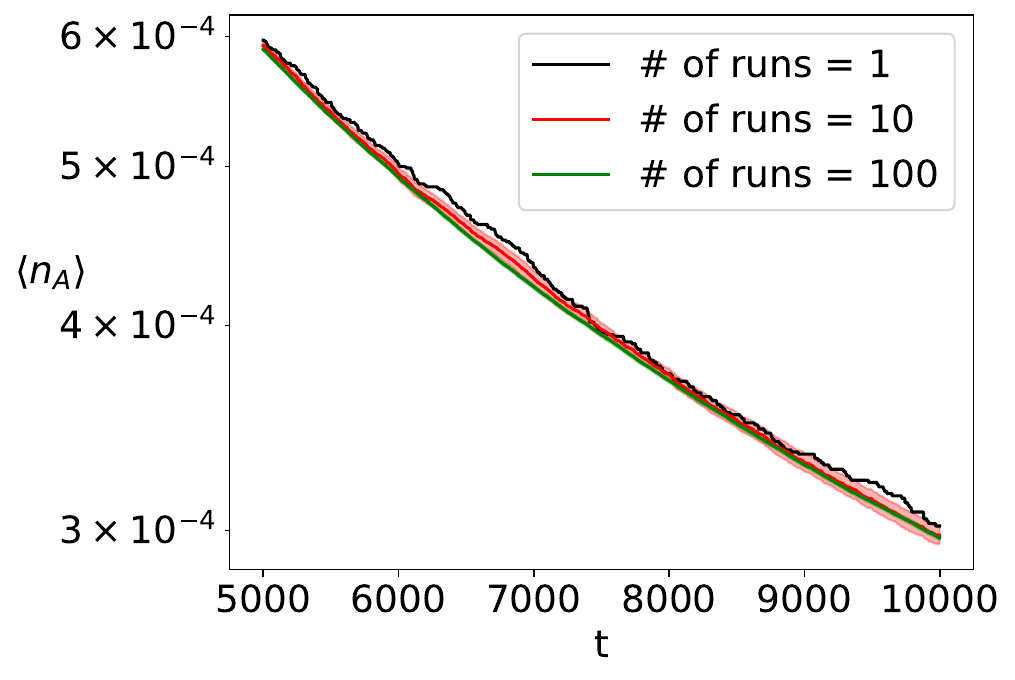}
    \caption{}
    \label{subfig:cog_multiple_runs}
    \end{subfigure}    
    \caption{(a) Individual stochastic realizations of the single-species coagulation model $2 \mathrm{A} \to \mathrm{A}$ in $d = 3$  shown in a lighter color, while the average over these trajectories is plotted in a thicker red line;  $L = 100$, $N_\mathrm{A}(0) = 10^6$, and $\lambda_0 = 1$.
	(b) Density decay of a single stochastic trajectory (black), and averaged over $10$ (red) and $100$ (green) runs.  
    The transparent fill region represents the standard error of the mean for the red line. 
    (It is not visible for the data in green).}
\end{figure}

We first revisit the density decay plots for the single-species coagulation model $2 \mathrm{A} \to \mathrm{A}$ discussed in Sec.~\ref{DLs1d}.
To visualize how distinct realizations determined by varying sequences of random numbers, it is common to plot the quantity of interest (here, the particle density) of each realization with a lower transparency value, and superimpose the resulting mean by a thicker line as shown in Fig.~\ref{subfig:cog_3d_runs_with_mean}. 
Such a plot visualizes the statistical spread of the data, and provides a useful representation of multiple trajectories in a single plot.
In this case, the stochastic deviations are small compared to the mean, and hence a small number of realizations is sufficient to accurately capture the decay.
This conclusion is further supported by Fig.~\ref{subfig:cog_multiple_runs}, which displays the means over different numbers of realizations.
The decay slope, even for a single run, is roughly captured despite the non-smooth trajectory.
The mean of a mere $10$ realizations becomes almost indistinguishable from the mean over $100$ trajectories, indicating that for the density decay, a few tens of independent runs suffice.

The main advantage of increasing the number of realizations is the ensuing reduction in the standard error of the mean. 
The standard deviation (the square root of the variance) $\sigma$ quantitatively describes the fluctuations relative to the mean of an observable (see Fig.~\ref{subfig:cog_3d_runs_with_mean}); $\sigma$ quantifies the amount of  noise in the system. 
In thermal equilibrium, the standard deviation of an observable is proportional to the temperature, permitting its  physical interpretation.
In contrast, the standard error of the mean (SEM) is useful for comparing results to other simulations and experiments.
For example, if we    compute the mean over $N$ realizations, the standard error of the mean describes how different the mean would be if another (or same) researcher did   the simulations again and computed their respective averages.
If we assume that these repeated  runs are statistically independent, we may use the central limit theorem and show that in the limit $N \to \infty$, the probability distribution of the mean of the density data is a Gaussian  with standard error $\sigma_\text{SEM} = \sigma / \sqrt{N}$.

\begin{figure}[t]
	\begin{subfigure}[t]{0.48\textwidth}
    \includegraphics[width=\linewidth]{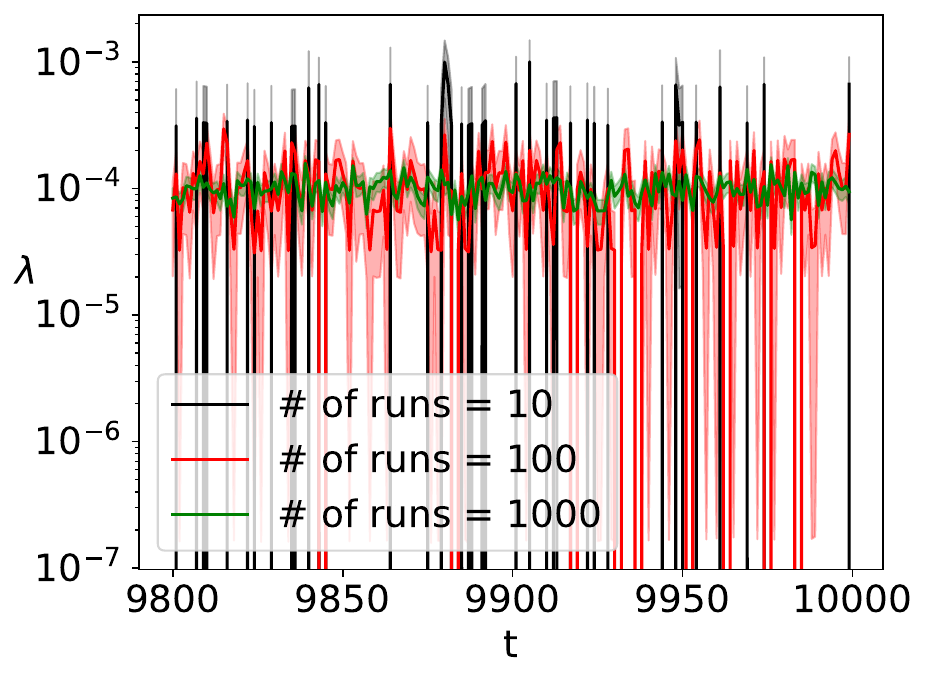}
    \caption{}
    \label{subfig:cog_rate_vs_runs}
    \end{subfigure}
	\hfill
	\begin{subfigure}[t]{0.48\textwidth}
    \includegraphics[width=\linewidth]{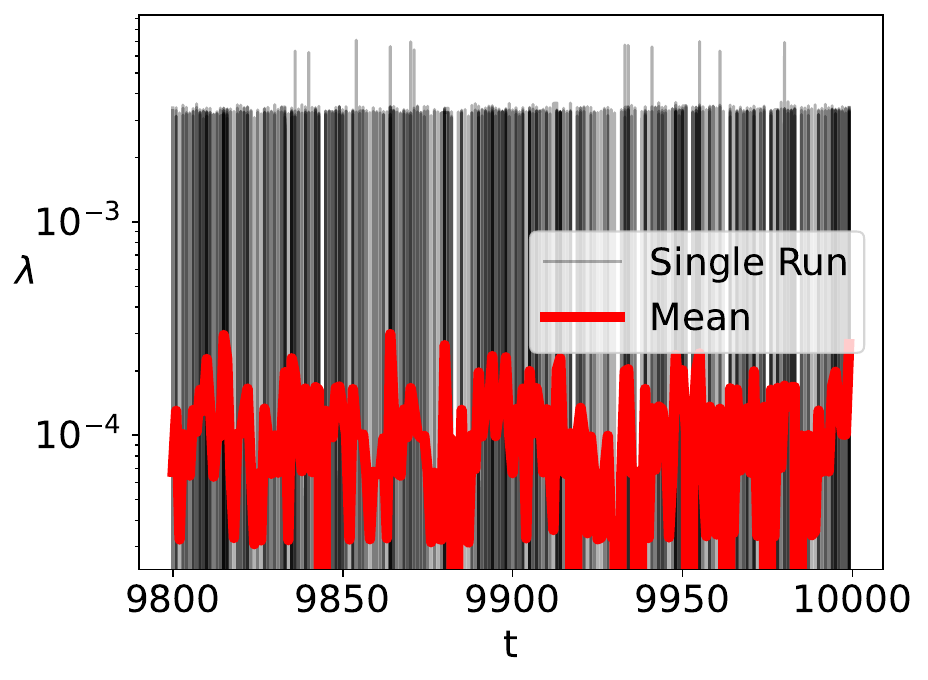}
    \caption{}
    \label{subfig:cog_rate_single_run}
    \end{subfigure}
    \caption{The effective coagulation rate for the binary reaction $2 \mathrm{A} \to \mathrm{A}$ in $d=3$  (parameters as in Fig.~12):
    (a) Averages over $10$ (black), $100$ (red), and $1000$ (green) realizations. 
    The transparent fill region represents the standard error of the mean for the red and green lines. 
    (b) Individual trajectories for the effective coagulation rate plotted in a lighter color, with their average shown in thicker red.}
    \label{fig:cog_runs_rate}
\end{figure}

Because only a relatively small number of independent realizations is required to obtain statistically significant data for the mean density decay in the single-species coagulation model, we might be tempted to conclude that the same is true for other quantities of interest.
That is not the case, as is demonstrated in Fig.~\ref{fig:cog_runs_rate} where a similar analysis is repeated for the effective coagulation rate (defined in Sec.~\ref{ere}). We see that that the fluctuations in the effective rate measurements are considerably larger than the spread of the density data.
Both plots indicate that as the density diminishes with time, coagulation events become increasingly rare, which amplifies their relative fluctuations.
Another contributing factor is that the effective coagulation is related to the derivative of the density, see Eq.~(\ref{eq:lambda_scaling_coagulation}), and generally, slopes of derivatives are very noisy.
We see that to obtain a statistically significant result for the effective coagulation rate, we need to average over $1000$ runs.

Satisfactory quantitative measurements of correlation functions typically also require many  runs.
Yet there are helpful methods to facilitate good statistics for correlations even for fewer runs. 
For example, in a stationary regime where time translational invariance holds, the correlation functions satisfy $C(\vec{x}, t; \vec{x}', t') = C(\vec{x}, t-t'; \vec{x}', 0)$.
Because only the time difference enters the correlations, we may obtain independent data beginning at different reference times $t'$.
Equal-time correlations are  independent of time and can simply be averaged over steady-state data if the reference time points are separated farther than the typical relaxation time.
Similarly, in systems with spatial translational symmetry, the correlation function depends only on position vector differences, and we may infer local correlations efficiently by averaging over sufficiently distant (beyond the characteristic correlation length) disjoint regions.
Care needs to be taken, because these symmetry properties usually hold for ensemble averages, and not for single runs.

Often the parameter region of interest is close to extinction for one of the species.
For example, annihilation reactions are characterized by universal scaling laws in the long-time limit when the absorbing (almost) empty state of the system is approached.
Another example is the directed-percolation critical point of the Lotka--Volterra predator-prey model, where scaling laws are observed close to the predator extinction threshold.
In such situations, the relative fluctuations become increasingly prominent as extinction is approached, which adversely affects the numerical determination of any observables of interest.
To obtain statistically significant results close to extinction regions, it is advisable to run the system on larger lattices, thus allowing the near-extinction region to be probed with a still sizable surviving number of particles.
These larger system sizes can be used in tandem with a very large number of independent runs to enhance the reliability of the measured scaling exponents in near-critical systems.

\begin{itemize}

    \item[] {\bf Problem 17}: Simulate the Lotka--Volterra predator-prey model with finite carrying capacity with different numbers of runs in the following parameter regions: (a) in the active coexistence regime; (b) in the predator extinction/prey fixation regime; and (c) in the prey fixation regime near the non-equilibrium critical transition point. 
    Compare the number of realizations needed to obtain sufficient statistics. 
    In particular, what happens near the predator extinction transition threshold?
    
\end{itemize}

\subsection {Finite-size effects and finite-time limitations}

Simulations are performed in finite systems, and often on lattices with a nonzero lattice spacing.
Similarly,  many experiments are conducted in constrained geometries, for example, biomolecular systems in a cellular environment.
In contrast, ecological systems and especially microbial systems are usually sufficiently large, 
so that their space is best considered infinite.
It might seem that the utility of simulations is restricted to finite-sized systems. This  limitation can be remedied by simulating sufficiently big systems and adequate extrapolations.
We briefly address how finite-size  affects the results of a simulation, and provide some analysis tools to differentiate between finite-size effects and  infinite-system behavior.

\begin{figure}[t]
	\includegraphics[width=0.48\textwidth]{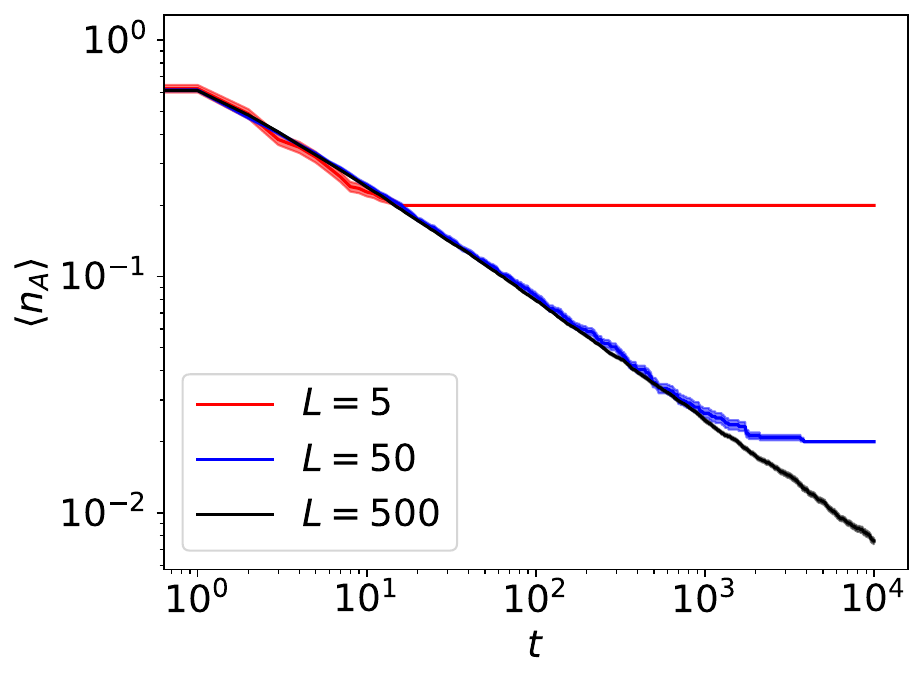}
	\caption{Particle density decay for the $d=1$ coagulation model $2 \mathrm{A} \to \mathrm{A}$  for various  lattice sizes and $N_\textrm{A}(0) = L$, $\lambda_0 = 1$.}
	\label{fig:finite_size_cog}
\end{figure}

Various quantities may depend crucially on the lattice size $L$. 
In particular, near critical points it also may modify the scaling exponents.\cite{Goldenfeld1992, Henkel2008, Tauber2014}
To illuminate the potential impact of varying $L$, we  consider the reactant density decay in the $d=1$ pair coagulation model $2 \mathrm{A} \to \mathrm{A}$.
Figure~\ref{fig:finite_size_cog} shows the data obtained in   simulations with increasing   $L$.
We see that the value of $L$ does not modify the long-time decay exponent.
However, the system  reaches its unique absorbing state with only one particle remaining at very early times for smaller lattices.
For $L = 5$, this discrepancy occurs at $t_L \approx 10$, and for $L = 50$ reactions still occur until $t_L \approx 1000$, consistent with characteristic diffusive scaling $t_L \sim L^2$. 
Beyond $t \geq t_L$, temporal measurements are affected by recurrent phenomena in finite systems.
Larger values of $L$  are required to unambiguously attain universal long-time scaling regimes and to reliably extract  exponents with adequate precision.

\begin{figure}[t]
	\includegraphics[width=0.48\textwidth]{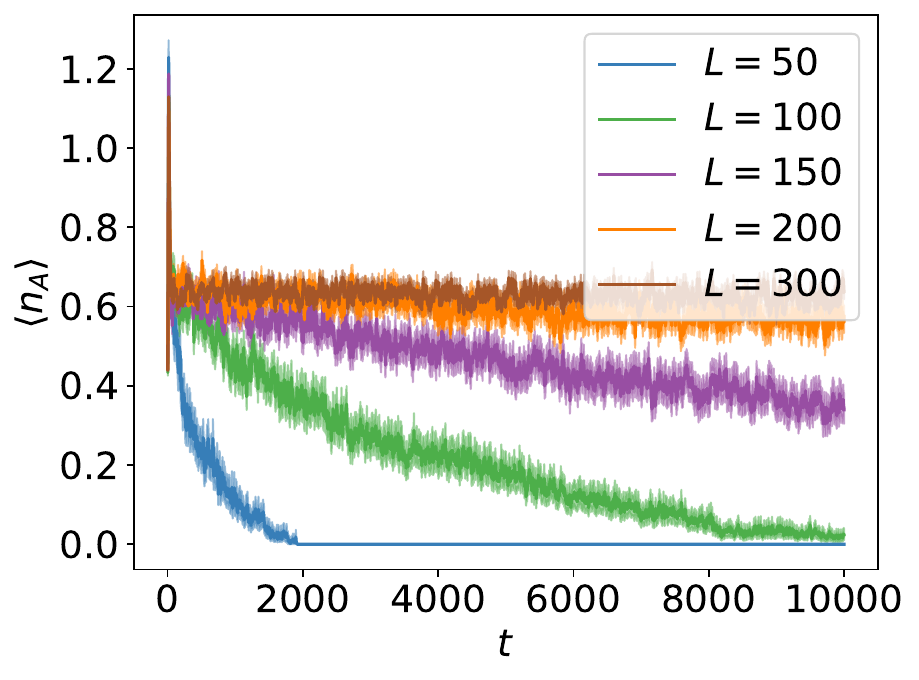}
	\caption{Predator density obtained for the $d=1$ Lotka--Volterra model  for various  values of $L$  and $n_A(0) = 0.5 = n_B(0)$, $\sigma_0 = 2$, $\mu_0 = 1$, $\lambda_0 = 1$, and carrying capacity $K = 5$. 
    For $L \to \infty$, this parameter set is in the two-species coexistence phase.}
	\label{fig:finite_size_LV}
\end{figure}

In the Lotka--Volterra predator-prey model, the species' densities can be drastically affected by finite-size effects, as demonstrated in Fig.~\ref{fig:finite_size_LV} for the predator population $\langle n_\mathrm{A}(t) \rangle$ in $d=1$ (with periodic boundary conditions).
For $L = 50$ and $100$, we observe the predator population go extinct, but attain a finite stationary density for larger lattices. 
All finite systems with absorbing states will ultimately go extinct because there must be a nonzero probability to reach extinction.
Usually, such extinction events, driven by fluctuations, become exceedingly rare in systems comprising many constituents, and correspondingly, mean extinction times typically grow exponentially with $L$ (see Problem~18).
Consequently,  the long-time behavior of small systems deviates drastically from that of large or infinite systems.
For intermediate values ($L = 150$ and $200$ in Fig.~\ref{fig:finite_size_LV}), we obtain a slowly decreasing mean density over time, which results from averaging the density time traces for different runs over multiple realizations, some of which reach the absorbing extinction state as $t$ progresses.

\begin{figure}[t]
\begin{subfigure}[t]{0.48\textwidth}
    \includegraphics[width=\linewidth]{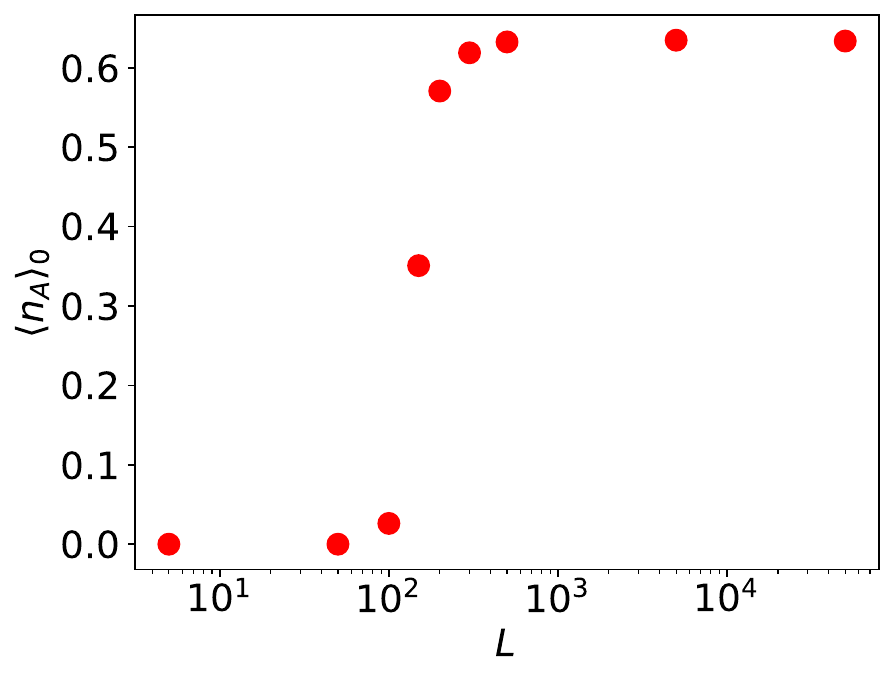}
    \caption{}
\label{subfig:finite_size_LV_nA}
\end{subfigure}
\hfill
\begin{subfigure}[t]{0.48\textwidth}
    \includegraphics[width=\linewidth]{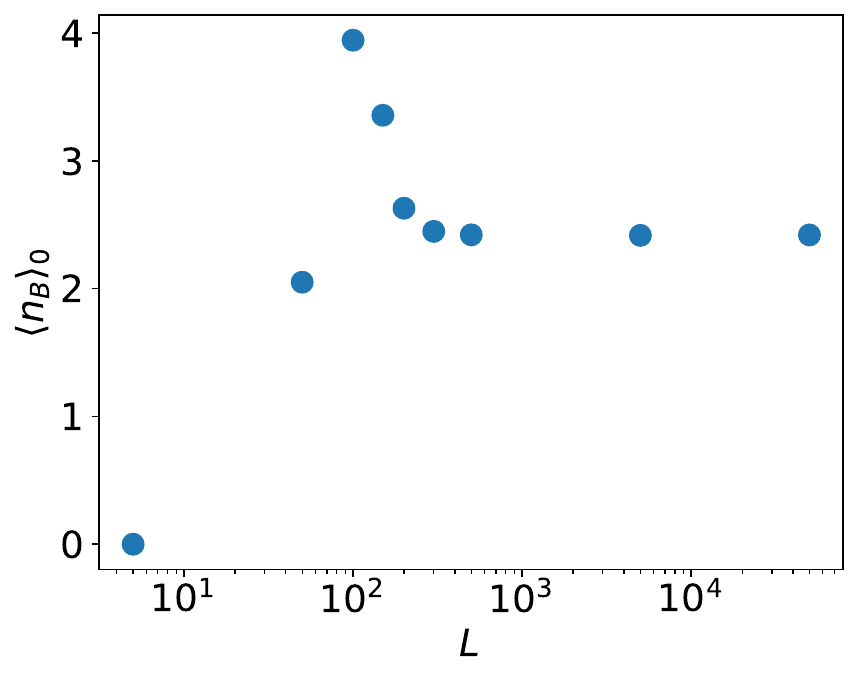}
    \caption{}
\label{subfig:finite_size_LV_nB}
\end{subfigure}
	\caption{Stationary (a) predator and (b) prey densities for the $d=1$ Lotka--Volterra model (parameters as in Fig.~\ref{fig:finite_size_LV}) as functions of the lattice size $L$.}
	\label{fig:finite_size_LV_stationary}
\end{figure}

This assertion is further supported by Fig.~\ref{fig:finite_size_LV_stationary}, which depicts the stationary predator and prey densities, computed by averaging over the last $10^3$ time steps, as a function of $L$.
The predator density exhibits a sharp transition from zero to size-independent finite values at a threshold $L_c \approx 150$, where the system becomes sufficiently extended and the particle number sufficiently large that none of the runs reach extinction within the simulation time.
Once $L$ is large enough, the stationary density becomes independent of $L$, indicating that this stationary density value can be utilized as a proxy for the $L=\infty$ limit.
The prey density correspondingly exhibits a cusp at $L_c$. 
For very small lattices, the prey species can also go extinct, which occurs for fewer runs as $L$ increases, because the predators are more likely to reach their absorbing state, so that the stationary prey population grows toward its maximum value equal to the carrying capacity $K$.
In the data shown in Fig.~\ref{fig:finite_size_LV_stationary}(b), about one quarter of the runs terminated in total population extinction.
For $L > L_c$, the predator species survive and control the prey density, whose stationary value quickly approaches its $L=\infty$ limit.
Although finite-size effects can prominently influence systems with absorbing states, we may still infer the infinite-system behavior from   simulations provided the observed quantities reach system size-independent values.

\begin{itemize}

    \item[] {\bf Problem 18}: 
    Any finite stochastic system with an absorbing state (such as species extinction) will eventually reach that absorbing state and stay there indefinitely. 
    However, it is often the case that the mean extinction time grows exponentially with system size. 
    Use the Lotka--Volterra model with $L < 100$ to confirm this statement. 
    To this end, perform many independent runs until each run reaches extinction at time $t_e$.
    Compare the averages $\langle t_e(L) \rangle$ for various $L$ and devise a plot to test the exponential dependence of the mean extinction time on $L$.
    
\end{itemize}

\subsection{Algorithmic artifacts}
\label{oalgoa}

As we have alluded, systems that exhibit multiple scaling regimes are prone to deviations from the asymptotic long-time, large-distance power laws if the corresponding prefactors differ drastically. 
For the pair coagulation model $2 \mathrm{A} \to \mathrm{A}$, the relevant microscopic time scales are set by the reaction rate $\lambda_0$ and the diffusivity $D_0$.
We have focused on the case where these two time scales are of the same order, $D_0 \approx \lambda_0 = \mathcal{O}(1)$.
We now fix the overall time scale by choosing $D_0 = 1$ and vary $\lambda_0$ to discuss how simulations in finite systems are affected when there is a large rate discrepancy $\lambda_0 \ll 1$ or $\lambda_0 \gg 1$.
For more complex systems, such time scale variations can strongly depend on the implementation of the Monte Carlo algorithm.
As is the case with finite-size effects, the results do not always constitute unwanted artifacts, but could also genuinely reflect model properties, depending on the system probed and the type of questions asked.

\begin{figure}[t]
\includegraphics[width=0.48\textwidth]{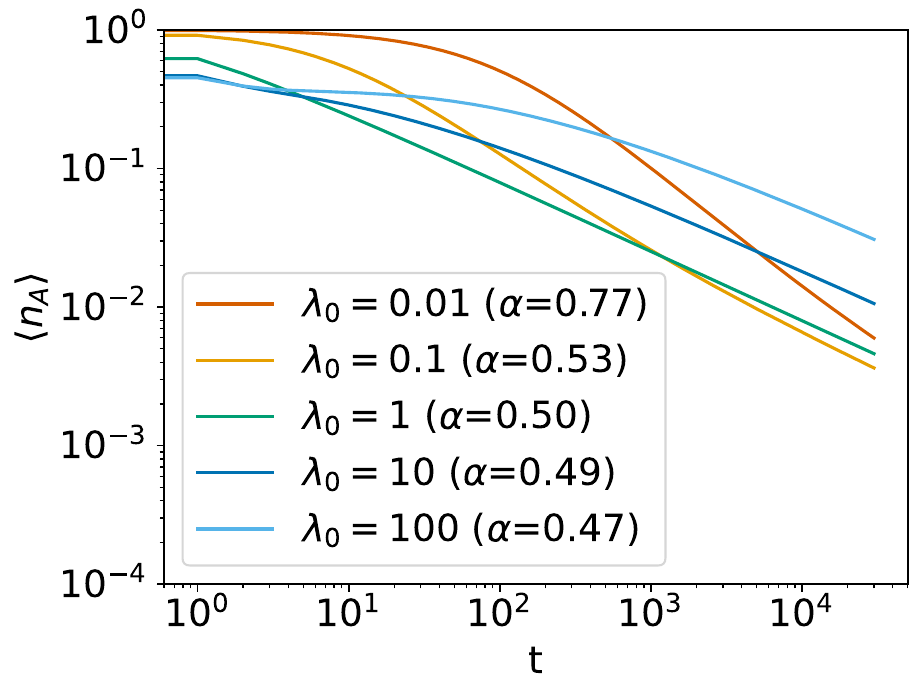}
	\caption{Particle density decay for $d=1$ coagulation $2 \mathrm{A} \to \mathrm{A}$ with various reaction rates $\lambda_0$ with $L = 1.5 \times 10^6$.
    The data were averaged over $10^3$ independent simulations. Power law fits were applied only to the final $10^4$ time steps.}
\label{fig:cog_timescale}
\end{figure}

Figure~\ref{fig:cog_timescale} shows the particle density decay exponent for different values of $\lambda_0$ for the same number of runs ($10^3$) and $L=1.5 \times 10^6$. The power law fit was computed using the last $10^4$ time steps.
Near $\lambda = 1$ the inferred decay exponent agrees with the theoretical value $\alpha = 0.5$. 
However, for $\lambda_0 \ll 1$,  the decay exponent appears to be larger than $0.5$, and seems slightly smaller than $0.5$ for $\lambda_0 \gg 1$.
The apparent decay exponent seems larger for smaller $\lambda_0$ because the system is still transitioning between  mean-field scaling with $\alpha = 1$ to diffusion-limited scaling with $\alpha = 0.5$, because for $\lambda_0 < 1$, the rate-limiting factor is the time it takes for a reaction to occur.
This problem can be remedied by increasing the simulation time and sampling more data points in the long-time limit.

This example showcases the effect of time scale separation, where arbitrary effective exponents between $0.5$ and $1$ may be observed depending on the sampled time window.
In contrast, if $\lambda_0 > 1$, diffusion is slow and the particles seldom interact, leading to a flatter density curve and thus explaining the smaller apparent decay exponent.
The artifacts produced by slower diffusion are less drastic than those generated by lower reaction rates, because even when $\lambda_0 > 1$, the system still ultimately resides in the diffusion scaling regime.
Because the slow diffusion features are due to the hopping reaction now becoming the rate-limiting rare event, this  deviation in the scaling exponent
can be resolved by sampling considerably more runs.
Consequently, for large rate discrepancies we would ultimately not observe distinct asymptotic scaling behavior, but would need to invest increased computational resources, such as running longer or doing more independent runs.

\begin{figure}[t]
\includegraphics[width=0.48\textwidth]{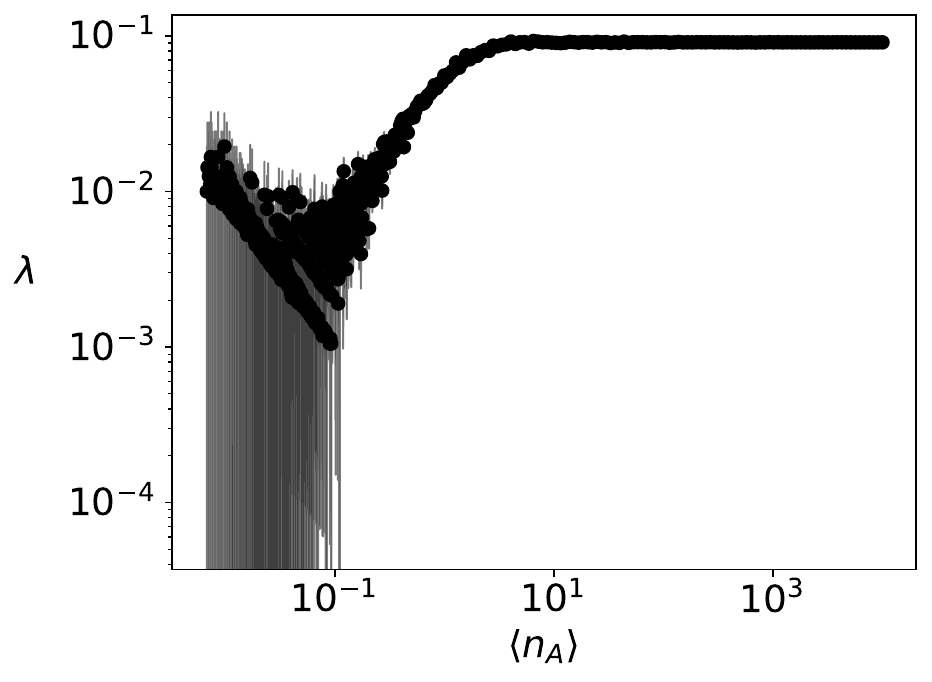}
	\caption{The effective reaction rate $\lambda$ versus the particle density $\langle n_\mathrm{A} \rangle$ for the $d=1$ coagulation model with multiple site occupancy ($L = 10^3$, $N_\mathrm{A}(0) = 10^4$, $\lambda_0 = 0.1$). 
    The shaded region shows the standard error of the mean. 
    The data were averaged over $10$  runs.}
    \label{fig:cog_multiple_occupancy}
\end{figure}

In our discussion of  annihilation models in Sec.~\ref{dlannr}, only a single particle was allowed per lattice site corresponding to a local carrying capacity $K = 1$.
When multiple occupancies are permitted and the reactant density is high, each particle exists in a background sea of other particles.
Therefore, in this case each particle always has another reactant available either on site or in its immediate neighborhood to initiate a pair coagulation or annihilation reaction.
This availability effectively removes the binary reaction condition that two particles need to meet, and consequently this case can be effectively represented by the simple decay process $\mathrm{A} \xrightarrow{\lambda_0} \emptyset$.
The ensuing exponential decay dominates at early times, until eventually the particle density becomes sufficiently reduced that multiple site occupancies are unlikely. 
The subsequent long-time limit will still be described by the pair coagulation power law.
Figure~\ref{fig:cog_multiple_occupancy} shows the effective macroscopic reaction rate for the pair coagulation reaction for $d=1$ with multiple site occupations and large initial density $\langle n_\mathrm{A}(0) \rangle = 10$.
The effective reaction rate $\lambda$,  Eq.~(\ref{eq:lambda_scaling_coagulation}), is initially constant and approximately equal to $\lambda_0$ as shown in Fig.~\ref{fig:cog_multiple_occupancy} for large $\langle n_\mathrm{A} \rangle$, as we  expect from the spontaneous decay process $\mathrm{A} \xrightarrow{\lambda_0} \emptyset$.
As the density decreases to $\mathcal{O}(1)$, the dependence of $\lambda$ on $\langle n_\mathrm{A} \rangle$ transitions to the diffusion-limited scaling behavior of the coagulation model with $\theta = 2$ as in Fig.~\ref{fig:coagulation_effective_rate}.
This analysis demonstrates that multiple site occupancies at high densities may lead to qualitatively different behavior whose  features are not incorrect, but do not reflect the desired model properties. 
It is therefore important to  understand the effective models that emerge as a result of different algorithmic and system choices.

\section{Conclusion}
\label{concls}

We have discussed  agent-based simulations of stochastic models that describe bio-chemical reactions, population dynamics, and epidemic spreading. We have specifically discussed potential pitfalls not usually mentioned in the literature related to obtaining adequate statistics, reaching desired scaling limits, and properly interpreting data for  the invariably simplified and effective  models that are intended to represent much more complex systems in nature.

We have emphasized the broad applicability of these comparatively simple models. 
It is straightforward to devise many varieties of reaction-diffusion models.
It is less obvious if and how these simple models are related and applicable to the real world.
Naturally, the underlying assumption is that much of a complex system's features are irrelevant in capturing the essence of its dynamics, or at least its gross dynamical properties of interest.
The art of devising good representations of nature consists of focusing on the truly important ingredients and omitting less crucial details.
If necessary, at least some of these additional complications can be reintroduced at the expense of more parameters and considerably increased computational costs.
Thus, we end with a final cautionary statement:
Although it is conceptually easy to add more variables and additional dynamical features to a model system, it becomes quickly impractical to properly run parameter sweeps to adequately assess their characteristic regimes.
This limitation ultimately constrains the applicability of stochastic models and the utility of the simulation tools employed for their investigation.

\section*{Author declarations}

\leftline{\bf Conflict of interest}

The authors declare no conflicts of interest. 

\leftline{\bf Data availability}

The original data used in the figures are contained in the electronic figure files, and are also available upon reasonable request to the authors.
Similar data can be generated using the codes in the browser-based Python Notebook.\cite{github2} 

\leftline{\bf Author contributions}

All authors contributed to writing the original manuscript, with final edits implemented by U.C.T. and the journal editors.
M.S., U.D., and R.I.M. designed the figures.
U.D. generated the browser-based Python Notebook.\cite{github2}

\begin{acknowledgments}
We gratefully acknowledge our many past and present collaborators whose invaluable contributions have not only decisively influenced our original findings, but also crucially informed our current conceptual and technical understanding.
Part of the research reported here was supported by the U.S.\ National Science Foundation, Division of Mathematical Sciences under Award No.\ NSF DMS-2128587.
\end{acknowledgments}

\end{document}